\documentclass[journal=jacsat,manuscript=article]{achemso}
\usepackage{xcolor}
\usepackage{graphicx}
\usepackage{graphics}
\usepackage{scalefnt}
\usepackage[version=3]{mhchem} 

\author{Bruno H. S. Mendon\c{c}a}
\affiliation[UFMG]{Departamento de F{\'i}sica, ICEX, Universidade Federal de Minas Gerais, CP 702, Belo Horizonte 30123-970, MG, Brazil}
\email{brunnohennrique13@gmail.com}

\author{Elizane E. de Moraes}
\affiliation[UFBA]{Instituto de F{\'i}sica, Universidade Federal da Bahia, Campus Universit{\'a}rio de Ondina, Salvador 40210-340, BA, Brazil}

\author{Ronaldo J. C. Batista}
\affiliation[UFOP]{Departamento de F{\'i}sica, Universidade Federal de Ouro Preto, Campus Morro do Cruzeiro, Ouro Preto 35400-000, MG, Brazil}

\author{Alan B. de Oliveira}
\affiliation[UFOP]{Departamento de F{\'i}sica, Universidade Federal de Ouro Preto, Campus Morro do Cruzeiro, Ouro Preto 35400-000, MG, Brazil}

\author{Marcia C. Barbosa}
\affiliation[UFRGS]{Instituto de F{\'i}sica, Universidade Federal do Rio Grande do Sul, Porto Alegre 91501-970, RS, Brazil}

\author{H{\'e}lio Chacham}
\affiliation[UFMG]{Departamento de F{\'i}sica, ICEX, Universidade Federal de Minas Gerais, CP 702, Belo Horizonte 30123-970, MG, Brazil}

\title[An \textsf{achemso} demo]
  {Water diffusion in carbon nanotubes for rigid and flexible models}

\abbreviations{IR,NMR,UV}

\begin{document}

\begin{abstract}
We compared the diffusion of water confined in armchair and zigzag carbon nanotubes for rigid and flexible water models. Using one rigid model, TIP4P/2005, and two flexible models, SPC/Fw and SPC/FH, we found that the number of the number of hydrogen bonds that water forms depends on the structure of the nanotube, directly affecting the diffusion of water. The simulation results reveal that due to the hydrophobic nature of carbon nanotubes and the degrees of freedom imposed by the water force fields, water molecules tend to avoid the surface of the carbon nanotube. This junction of variables plays a central role in the diffusion of water, mainly in narrow and/or deformed nanotubes, governing the mobility of confined water in a non-trivial way, where the greater the degree of freedom of the water force field, the smaller it will be mobility in confinement, as we limit the competition between area/volume, and it no longer plays the unique role in changing water diffusivity.
\end{abstract}

\section{Introduction}
\label{sec:introduction}

Water permeates almost everything in Nature. It is present in all living organisms. It is hard to find a system which is not in contact with this curious substance directly or indirectly. Water can be found as vapor, liquid, or solid since its triple point lies in a range of temperatures and pressures naturally present on Earth's surface. Water also presents itself in any length scale -- from thousands of kilometers in bulky oceans down to a few nanometers as inside carbon nanotubes. Undoubtedly, any attempt to understand how this planet works includes the study of water and how their properties interfere in virtually every single aspect of our lives. 

Most physical quantities are affected by environment variables, specially those related to length scales. Macroscopic systems not necessarily work 
the same way they do at the nanoscale level. For example, measurements and computational simulations showed that water confined in carbon nanotubes (CNTs) is expected to have structural, transport, and dynamical properties different from those observed in the bulk liquid~\cite{@nature/35102535,@10.1021/nl048876h,@10.1038/43844a,@10.1021/nl052254u,@10.1021/nl080705f,@10.1007/s10404-010-0678-0,@10.1038/nature19315,@10.48550/arXiv.2001.00614,@10.1016/j.physa.2018.11.042,@10.1063/1.5129394,@10.1016/j.chemphys.2020.110849,@10.1021/acsnano.2c02784}. 

The water behavior in the nanoscale realm is of crucial interest to all scientific community, specially those related to nanofluidic properties, due to its importance in nonporous systems with a wide range of applications
based on nanotube membranes. Examples include water treatment technologies, energy storage systems, nanosyringes, drug delivery, intracellular solute transport, and cancer therapy~\cite{@10.1016/j.jmr.2014.07.012,@10.1021/nl048876h,@10.1126/science.1181799,@10.1039/B909366B,@10.1073/pnas.1004714107,@10.1126/science.1200488,@10.1186/1556-276X-6-555,@10.1038/nature11477,@10.1038/NNANO.2015.37,@10.1039/C3CS60253B,@10.1016/j.msec.2016.12.058,@10.1038/s41563-020-0625-8,@10.1038/s41563-020-00849-5}.

Diffusion is a crucial transport property reflecting the dynamic behavior of fluids. Water in the bulk phase presents Fickian diffusion, when the mean square displacement (MSD) is linear with time~\cite{@10.1021/nl052254u,@10.1088/0957-4484/18/47/475704}. 

When confined in CNTs, water diffusivity depends on the physical characteristics of the confining tube~\cite{@10.1021/nl052254u,@10.1073/pnas.1108073108}. For example, if the confinement prevents the water molecules to pass each other, the diffusion can occur in a single-file mode, e.g. MSD $\sim t^{\frac{1}{2}}$~\cite{@10.1126/science.287.5453.625,@nature/35102535}. In the case where the water molecules move coordinated, their diffusion can occur in a ballistic mode, e.g. MSD $\sim t^{2}$~\cite{@10.1021/nl052254u,@10.1073/pnas.1108073108}. Furthermore, when the water molecules can pass through each other the diffusion is of the Fickian type as in the case of the bulk phase~\cite{@10.1063/1.2131070,@10.1063/1.2565806}. 

Theoretical studies using Molecular Dynamics (MD) simulations of water molecules confined in CNTs of different diameters observed that the diffusion coefficient of the confined water is non-monotonically as a function of the diameter, which can be ascribed to the surface effect and the size effect of CNTs~\cite{@10.1007/s10404-011-0772-y,@10.1021/jp205877b,@10.1016/j.physa.2018.11.042}. In addition, the faster diffusivity of water in CNTs could be attributed to the ordered hydrogen bonds formed between water molecules within the confined channels of CNTs and the weak interaction between water and the CNTs. In addition to this already complex scenario, the diffusion coefficient obtained by MD simulations depends on the details of the models used.

Water is a complex substance to model, because of the competing eﬀects of hydrogen bonding and van der Waals interactions. In the literature, there are several models to describe water~\cite{@10.1007/978-94-015-7658-1_21,@10.1021/jp003919d,@10.1063/1.1683075,@10.1063/1.445869,@10.1063/1.481505,@10.1063/1.2056539,@10.1063/1.2907845,@10.1063/1.3124184,@10.1103/PhysRevLett.126.236001,@10.1080/00268979100102391,@10.1039/D0NR02511A}. These models are constructed to fit a set of experimental data, and their success depends on being able to reproduce additional experimental properties both in the bulk and in confined water. 

TIP4P/2005 is one of the most used rigid model~\cite{@10.1063/1.2121687}. It is composed by four points: the oxygen with mass, two hydrogen with positive charges and a fictitious location between the oxygen and the hydrogen to represent the dislocated charge of the oxygen. This model was parameterized using experimental data such as the maximum density temperature, the enthalpy of vaporization, and the density of liquid water at ambient conditions. It can reproduce thermodynamic properties of water in a wide range of temperatures~\cite{@10.1063/1.2121687} but is not able to reproduce processes involving chemical bond formation and dissociation.

In order to circumvent this limitation, flexible models were created. They represent the  O-H bond lengths and angles  by harmonic functions and are better equipped to reproduce flexibility  transport properties~\cite{@10.1080/00268978700100141}. The flexible models, SPC/Fw~\cite{@10.1063/1.2136877} and SPC/FH~\cite{@10.1063/1.3124184} models, were fitted from the SPC to describe dynamic properties.

From the experimental point of view, a great number of methods, such as infrared spectroscopy~\cite{@10.1126/science.282.5386.95,@10.1021/jacs.6b02635}, Raman spectroscopy~\cite{@10.1140/epjb/e2002-00224-8}, neutron scattering~\cite{@10.1063/1.2194020}, x-ray diffraction~\cite{@10.1143/JPSJ.71.2863,@10.1016/j.cplett.2004.11.112}, x-ray Compton scattering~\cite{@10.1103/PhysRevLett.111.036803}, and  nuclear magnetic resonance~\cite{@10.1103/PhysRevB.84.165417} are suitable to study the structure and dynamics of confined water inside CNTs. However, to the best of our knowledge, no experimental work fully explains how water organize and diffuse when confined. Interpreting experimental data is challenging because nanotubes are far from being perfect in real world (a fact that is commonly abstracted in theoretical models along with their internal degrees of freedom), presenting defects, vacancies and structural distortions~\cite{@10.1016/j.physa.2018.11.042,@10.1063/1.5129394} which may affect several of water properties~\cite{@10.1016/j.commatsci.2004.02.018,@10.1016/j.carbon.2015.09.099,@10.1016/j.physa.2018.11.042,@10.1063/1.5129394}. 

In this context, the main objective of this study is to determine how flexible water models differ from the rigid ones when simulating water diffusion in carbon nanotubes, and also how water molecules organize under severe confinement. We considered different nanotube chirality structures, in cases where tubes are perfect, kneaded, and wrinkled.

We examine the structural and dynamic properties of confined water affected  by chirality and nonuniform deformations in nanotubes. Furthermore, we considered perfect, kneaded, and wrinkled tubes with zigzag and armchair chirality types~\cite{@10.1063/1.5129394}.
Internal degrees of freedom of carbon nanotubes were modelled by Morse, harmonic, and Lennard-Jones potentials. For the water, we considered three simulation models. The rigid TIP4P/2005 model~\cite{@10.1063/1.2121687} and two flexible ones, which are SPC/E variants: the SPC/FH~\cite{@10.1063/1.3124184} and SPC/Fw~\cite{@10.1063/1.2136877}. The SPC/Fw and SPC/FH models, unlike the TIP4P/2005, allow the variation of both the angles and the equilibrium bonding distances of the water molecule. The influence of the introduction of these degrees of freedom in the system and the different parameters adopted in each force field, alter the structural and dynamic behavior of the water and our objective is to analyze this phenomenon when we confine the water molecules in carbon nanotubes with two types of deformation.

The remaining of this paper goes as follows. In Sec. II computational details are presented. Results are discussed in Sec. III, while conclusions are shown in Sec. IV.

\section{The Models and the Simulation Details}
\label{sec:models}

We performed molecular dynamics simulations with constant number of particles, volume, and temperature to analyze how different water models affect the results for the diffusion coefficient of confined water. For water, we used the rigid model TIP4P/2005~\cite{@10.1063/1.2121687} which reproduces several thermodynamic water properties conditions~\cite{@10.1039/B805531A,@10.1080/00268970902784926} and the flexible  SPC/Fw~\cite{@10.1063/1.2136877} and SPC/FH~\cite{@10.1063/1.3124184} models which reproduce water dynamic properties. 

The TIP4P/2005 represents water by a rigid four point structure where with one point at the oxygen which has no charge, two in the hydrogen and another point M where the oxygen charge is located. The water-water interaction is modeled by oxygen-oxygen Lennard-Jones plus coulomb for the hydrogen and M-point interactions. The parameters are shown in Table~\ref{tab_watermodels}. The SPC/Fw and SPC/FH models are inspired in the SPC rigid model, which has three points: one for the oxygen with Lennard-Jones and coulomb interactions and two  hydrogen with coulomb interactions. In addition to the Lennard-Jones and coulomb interactions, the flexible models allow the variation of both water equilibrium bond distances and angles with harmonic interaction terms. The SPC/Fw and the SPC/FH parameters are illustrated in Table~\ref{tab_watermodels}. The comparison between the models indicates that the SPC/FH is more rigid with larger spring constants than the SPC/Fw. The SPC/FH includes a Lennard-Jones hydrogen-hydrogen interaction, allowing a stronger interaction with neighboring molecules.

The study of the impact of introducing these degrees of freedom in the system, as well as the different parameters adopted in each force field, provides a necessary overview to determine the most suitable model to study the systems analyzed in this work.

\begin{figure}[H]
	\begin{center}
		\includegraphics[width=6.4in]{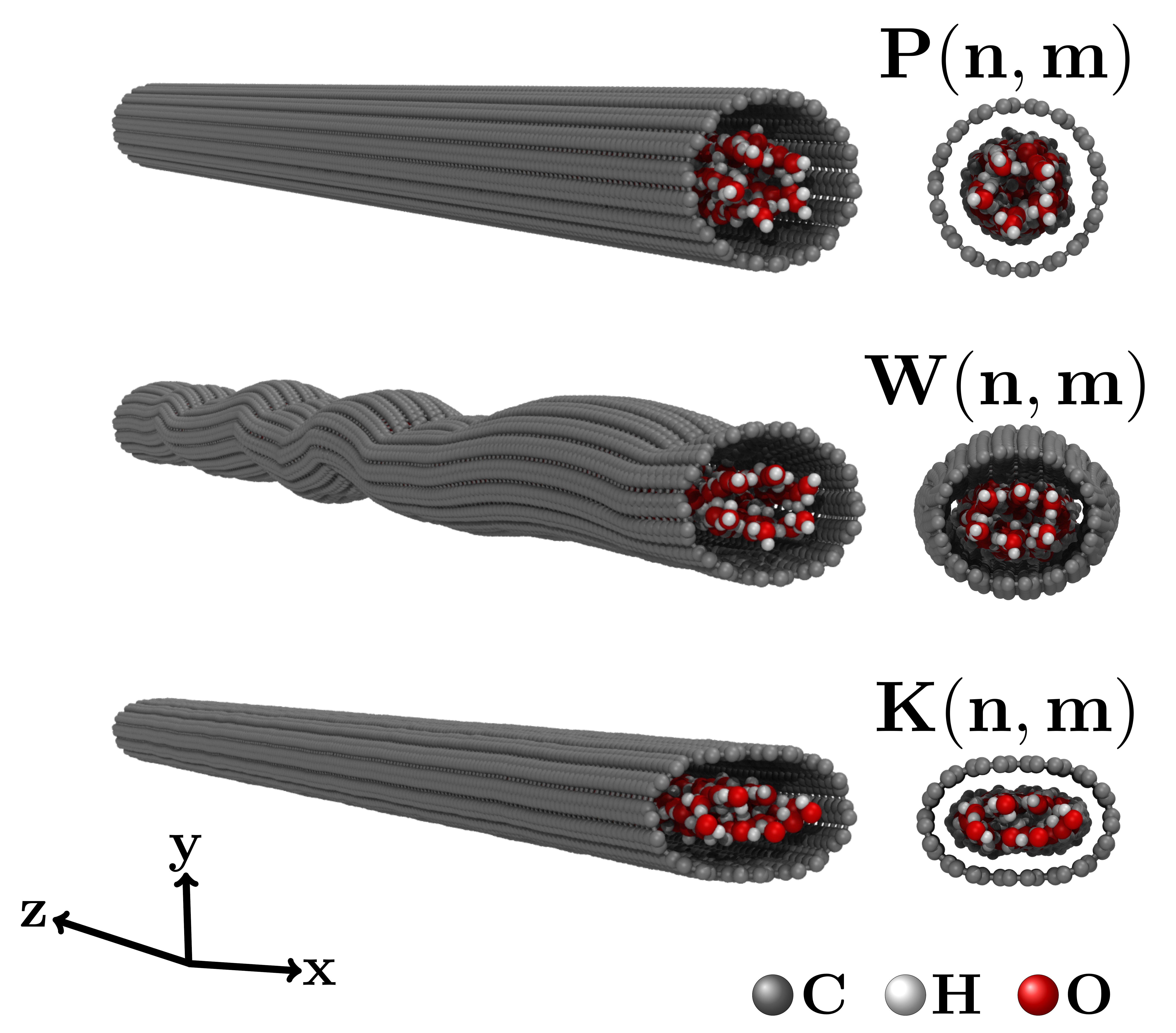}
	\end{center}
	\caption{Snapshot of the perfect P($n,m$), wrinkled W($n,m$) and kneaded K($n,m$) simulated carbon nanotubes.}     
	\label{fig_cnts}
\end{figure}

Water molecules were confined in carbon nanotubes (CNT) with different diameters, chirality, and two types of deformation: kneaded $K$ and wrinkled $W$ (See Figure~\ref{fig_cnts}). Following the notation ($n,m$) to characterize the chirality of CNTs, we use three armchair nanotubes ($n=m$), namely, ($7,7$), ($9,9$) and ($12,12 $) and three zigzag nanotubes ($m=0$), namely ($12,0$), ($16,0$) and ($21,0$). The diameter of the CNTs can be given as a function of the indices $n$ and $m$ as $d= (\sqrt{3}/\pi)a\sqrt{n^2 +m^2 + nm}$, where $a=1.42$~\AA\ is the C-C bond length.

In order to investigate the effects of radial asymmetry on the diffusion of confined water, we uniformly deformed the nanotubes to different degrees. The kneaded nanotubes were produced by taking a perfect nanotube and uniformly kneaded it in the $y$ direction until the nanotube reached an elliptical cylindrical shape with eccentricity equals to $0.8$. We define the degree of deformation by the eccentricity of the ellipse being $e=\sqrt{1 - \alpha^2/\beta^2}$, where $\alpha$ and $\beta$ are the semi-minor and semi-major axis of tube's straight section, respectively. The wrinkled nanotubes were produced by randomly compressing the tube in the $y$ direction until non-uniform wrinkles are formed. Wrinkles were created in a disorderly fashion, but as the number of wrinkles is small, the size distribution of the segments between two wrinkles were the same for nanotubes of the same diameter. On average, each nanotube was compressed at five different positions on the $z$ axis, with eccentricity values ranging between $0.0$ and $0.8$. 

Furthermore, for comparison purposes, perfectly structured nanotubes were also produced. Perfectly symmetrical nanotubes are characterized by an eccentricity of $0.0$. We defined the three types of nanotubes used, both armchairs and zigzags, as perfect P($n,m$), kneaded K($n,m$) and wrinkled W($n,m$).

The carbon-carbon and carbon-water interaction were modelled using the Lennard-Jones potential (LJ)~\cite{@10.1088/0959-5309}. The classical potential for the interaction between carbon atoms has been described with an energy of $\epsilon_{CC}=0.086$~kcal$\cdot$mol$^{-1}$ and an effective diameter of $\sigma_{CC}=3.4$~\AA~\cite{@nature/35102535}. The carbon-oxygen energy $\epsilon_{CO}=0.11831$~kcal$\cdot$mol$^{-1}$ and the effective carbon-oxygen diameter $\sigma_{CO}=3.28218$~\AA~\cite{@nature/35102535}. The parameters considered for the water models were defined and are shown in the Table~\ref{tab_watermodels}.

\begin{table}[H]
	\begin{center}
	\caption{Force field parameters used for each of the water models. The Lennard-Jones site is located on the oxygen atom, with parameters $\sigma$ and $\epsilon$. The charges of oxygen and hydrogen are $q_{O}$ and $q_{H}$, respectively. The TIP4P/2005~\cite{@10.1063/1.2121687} model places a negative charge $q_{M}$ at a point M at a distance $d_{OM}$ from the oxygen along the H-O-H bisector. The distance between the oxygen and hydrogen sites is $r_{OH}$. The angle formed between hydrogen, oxygen and another hydrogen atom is given by $\theta_{HOH}$. For flexible models (SPC/Fw~\cite{@10.1063/1.2136877} and SPC/FH~\cite{@10.1063/1.3124184}), the $k_{OH}$ and $k_{\theta}$ are the potential depth parameters, and OH and $\theta$ are the reference bond length and angle, respectively.}
		\begin{tabular}{ c c c c }
			\hline
			  & TIP4P/2005 & SPC/Fw & SPC/FH   \\ \hline
                $\epsilon_{OO}$ (kcal mol$^{-1}$) & 0.1852 & 0.155 & 0.1553 \\
                $\epsilon_{HH}$ (kcal mol$^{-1}$) & 0.0  & 0.0 & 0.0396 \\
                $\sigma_{OO}$ (\AA) & 3.1589 & 3.165 & 3.188 \\
                $\sigma_{HH}$ (\AA) & 0.0 & 0.0 & 0.65 \\	
                $q_{O}$ (e) & 0.0 & -0.82 & -0.8476 \\	
                $q_{H}$ (e) & 0.5564 & 0.41 & 0.4138 \\	
                $q_{M}$ (e) & -1.1128 & * & * \\
                $d_{OM}$ (\AA) & 0.1546 & * & * \\	
                $r_{OH}$ (\AA) & 0.9572 & 1.012 & 1.0 \\		
                $\theta_{HOH}$ ($^{\circ}$) & 104.52 & 113.24 & 109.4 \\
                $k_{OH}$ (kcal mol$^{-1}$ \AA$^{-2}$) & * & 1059.162 & 1108.580 \\	
                $k_{\theta}$ (kcal mol$^{-1}$ rad$^{-2}$) & * & 75.90 & 91.53 \\	      	
                \hline 
		\end{tabular}
		\label{tab_watermodels}
	\end{center}
\end{table}

The density of water was determined considering the volume excluded due to the LJ interaction between carbon and oxygen atoms. Thus, the density is determined by $\rho=4M/\pi(d_{t}-\sigma_{CO})^2L_z$, where $M$ is the total mass of water in the tubes and $ L_{z}$ is the length of the nanotube. The CNT diameters $d_t$ range from $0.94$~nm to $1.64$~nm, the lengths L$_{z}$ range from $22.5$~nm to $123.4$~nm and the amount of water confined in each nanotube varies from 901 to 908 molecules, as shown in Table~\ref{tab_CNTp}. Deformed nanotubes were simulated with the same length as well as the same total mass of confined water of their the equivalent, perfect nanotubes.

\begin{table}[H]
	\begin{center}
	\caption{Parameters for the perfect carbon nanotubes. The nominal diameter $d_{t}$, the length L$_{z}$, the density $\rho$ and the amount of water molecules confined in each system.}
		\begin{tabular}{ c c c c c }
			\hline
			CNT & $d_{t}$ (nm) & L$_{z}$ (nm) & $\rho$ (g/cm$^{3}$) & H$_{2}$O \\ \hline
			(7,7) & 0.95 & 123.4 & 0.90 & 901 \\
			(12,0) & 0.94 & 123.0 & 0.91 & 901 \\
			(9,9) & 1.22 & 50.5 & 0.92 & 908 \\
			(16,0) & 1.25 & 50.5 & 0.80 & 908 \\
			(12,12) & 1.63 & 22.5 & 0.94 & 901 \\
			(21,0) & 1.64 & 22.9 & 0.86 & 901 \\ 
			\hline 
		\end{tabular}
		\label{tab_CNTp}
	\end{center}
\end{table}

Simulations were performed with the Large-scale Atomic/Molecular Massively Parallel Simulator (LAMMPS) package~\cite{LAMMPS,@10.1006/jcph.1995.1039}. We employed the Particle-Particle Particle-Mesh (PPPM) method to calculate long-range Coulomb interactions. This method handles the long-range interactions and the Coulomb field of real charges in a way that could interfere with their own images~\cite{@10.1021/acs.jpcc.7b08326}. We got around this problem by creating a $xy$ simulation box around $100$~nm for all the nanotubes, preventing the carbon nanotube from interacting with its own images and avoiding the overlapping of virtual images with real images, minimizing possible errors in the application of the method. 

Water temperature was maintained at $300$~K through a Nos{\'e}-Hoover thermostat with a damping time of $100$~fs and a time step of $1$~fs for the TIP4P/2005 model and a time step of $0.5$~fs for the SPC/Fw and SPC/FH models. In all simulations, we kept the nanotubes rigid  without out-of-plane displacements. This procedure was employed in several similar simulations, which showed that considering the nanotube as a rigid system is a very reasonable approximation when compared to the case where the thermostat is applied throughout the system~\cite{@10.1021/jp205877b,@10.1063/1.2187971,@10.1016/j.ijmultiphaseflow.2004.03.009}. All systems for TIP4P/2005 were balanced to $5$~ns, properties were stored every $0.01$~ns for $5$~ns, giving a total simulation time of $10$~ns. For the SPC/Fw and SPC/FH models all systems were balanced to $15$~ns, properties were stored every $0.01$~ns for $15$~ns, giving a total simulation time of $30$~ns.

Due to the dimensions of the system, diffusion is minimal in the radial direction and only axial diffusion is considered. Diffusion is determined using the one-dimensional Einstein relation:

\begin{equation}
    \label{eq:MSD}
    D_z = \lim_{\tau \to \infty} \frac{1}{2}\frac{d}{d\tau}\left\langle z^2(\tau)\right\rangle \;,
\end{equation}

\noindent where $\left\langle z^2(\tau)\right.\rangle=\left\langle\left[z(\tau_0-\tau)-z(\tau_0)\right]^2\right\rangle$ is the mean-squared displacement (MSD) in the axial direction of the system.

For characterizing the structure of water, we calculated the number of hydrogen bonds (HB) and also made density maps for the occurrence of oxygen in the $xy$ plane. The hydrogen bonds were calculated if both of the following geometric criteria were satisfied $\alpha\leq 30^{\circ}$ and $|\vec{r}_{OO}|\leq 3.50 \mbox{~\AA}$, where $\alpha$ is the angle $OH\cdots O$ and $|\vec{r}_{OO}|$ is the distance between two oxygen~\cite{@10.1021/nl072385q}. Oxygen occurrence density maps were obtained by dividing the corresponding plane $xy$ into square boxes $0.1$~\AA\ in length and counting the number of oxygen in each box. Higher oxygen densities are represented in red, while low densities tend to darker blue tones. On average, each simulation step consisted of three sets of simulations with different initial thermal velocity distributions.

\section{Results and discussion}
\label{sec:results}

Thermodynamic and dynamic quantities of  bulk water when modeled by the TIP4P/2005 and the SPC/Fw models have shown good  agreement with experimental results~\cite{@10.1063/1.2121687,@10.1080/00268970902784926,@10.1039/B805531A,@10.1063/1.2136877}. The SPC/FH also shows an ability to validate studies carried out, comparing with experimental result~\cite{@10.1063/1.3124184}. 

First, we compared the experimental diffusion coefficient for bulk water with the results of molecular dynamics simulations for the  TIP4P/2005, SPC/Fw and SPC/FH models for $298.15$~K. The experimental diffusion coefficient of bulk water is about $23\times10^{-10}$ m$^2$/s and for the TIP4P/2005 model is $20.8\times10^{-10}$ m$^2$/s, which agrees with several molecular dynamics simulations using the same model~\cite{@10.1039/F19807600377,@10.1063/1.2121687}. For the flexible models (SPC/Fw and SPC/FH) the  diffusion coefficients found were $23.59\times10^{-10}$ m$^2$/s and $21.89\times10^{-10}$ m$^2$/s respectively.\cite{@10.1063/1.4749382} The  diffusion coefficient for the SPC/FH model, computed in this simulation, is lower than the values obtained by SPC/Fw. This is consistent with SPC/FH being more structured  than the SPC/Fw model. The SPC/Fw diffusion is closer to the experimental value.

\begin{figure}[H]
	\begin{center}
		\includegraphics[width=5.in]{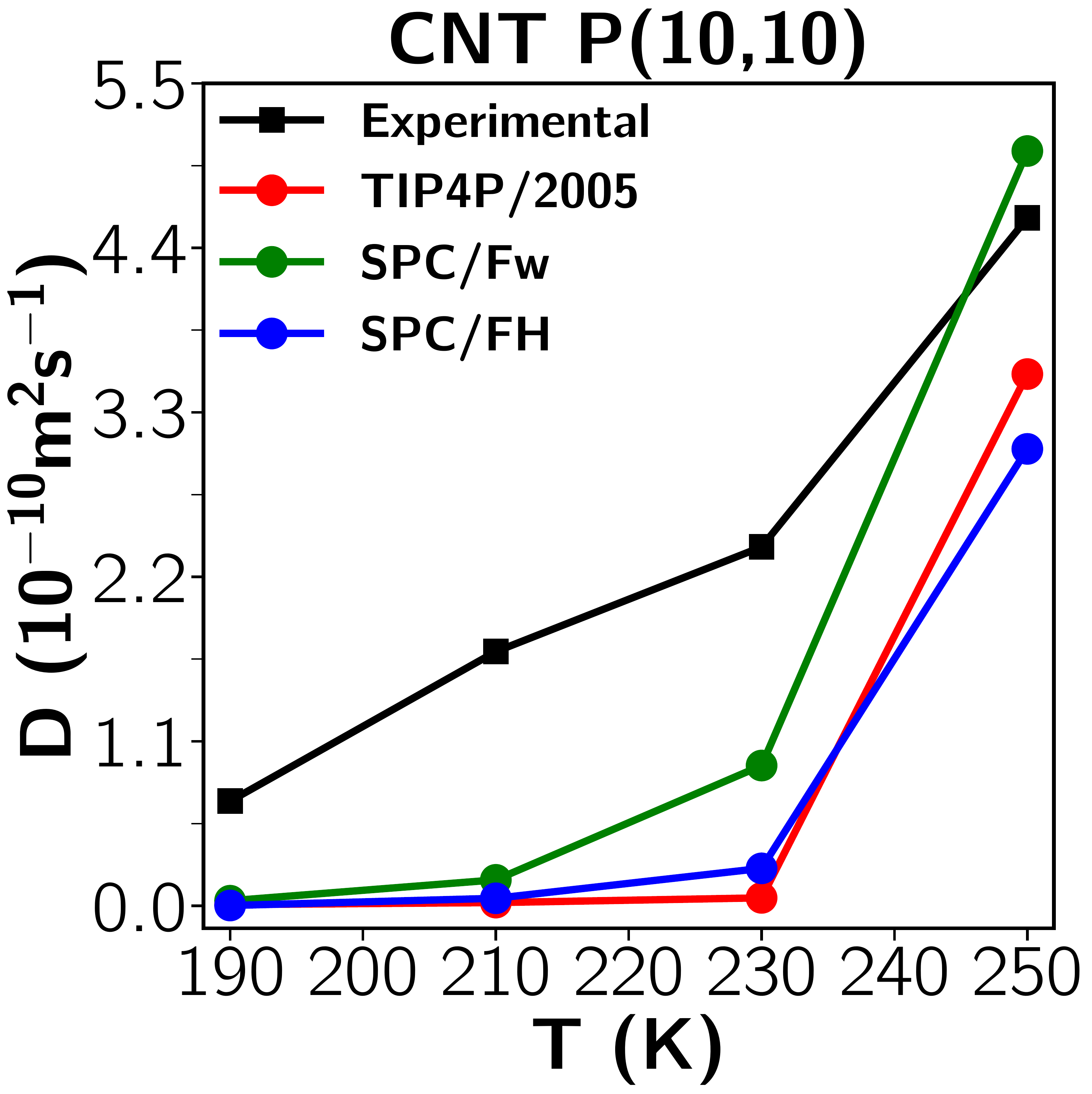}
	\end{center}
	\caption{Diffusion coefficient versus temperature for water  confined water in P($10,10$) carbon nanotubes for experimental-derived (black squares)~\cite{@10.1063/1.2194020} values, for simulations  using TIP4P/2005 (red circles), SPC/Fw (green circles) and SPC/FH (blue circles) water models.}
	\label{fig_diff-ref}
\end{figure}

Next, we compared the simulations with experimental results for the diffusion coefficient of confined water. Figure~\ref{fig_diff-ref} shows D versus temperature of water confined in a perfect armchair nanotube with size P(10,10). The black squares illustrate the experiment-derived estimates~\cite{@10.1063/1.2194020}, generated by a fitting function to experimental neutron scattering spectra. It is important to emphasize that, according to the authors, the temperature dependence of diffusivity is difficult to access experimentally. In the author's own words, the method ``may be unable to capture the entire range of diffusivity, thus missing a slower fraction and yielding an artificially enlarged effective diffusion coefficient''~\cite{@10.1063/1.2194020} and ``Because of the overestimate of the diffusion coefficient (and, therefore, mean jump length) is likely temperature dependent, the true temperature dependence of these parameters are difficult to assess.~\cite{@10.1063/1.2194020}. 

We calculated the diffusion coefficient of confined water in the same system and temperature range using the TIP4P/2005, SPC/Fw, and SPC/FH water models. The results of these calculations are also shown in Figure~\ref{fig_diff-ref} as red, green, and blue circles, respectively. The figure shows that, consistent with the comments in Ref.~\citenum{@10.1063/1.2194020}, in the $190-230$ K temperature range the experimentally estimated D values are considerably larger than all the three calculated values, indicating that these experimentally estimated values might indeed be artificially enlarged. Moreover, as also predicted in Ref.~\citenum{@10.1063/1.2194020}, the apparent overestimate of the experiment-derived values is strongly temperature-dependent, up to a point, at a temperature of $250$~K, where both calculated values and the experimental estimate of D have the same order of magnitude. 

The diffusion coefficients for the more structured systems, the rigid TIP4P/2005 and flexible SPC/FH models, are quite similar, while the SPC/Fw presents higher values that are closer to the experimental results and in confinement the flexibility of the SPC/Fw becomes even more relevant.

\begin{figure}[H]
	\begin{center}
		\includegraphics[width=6.4in]{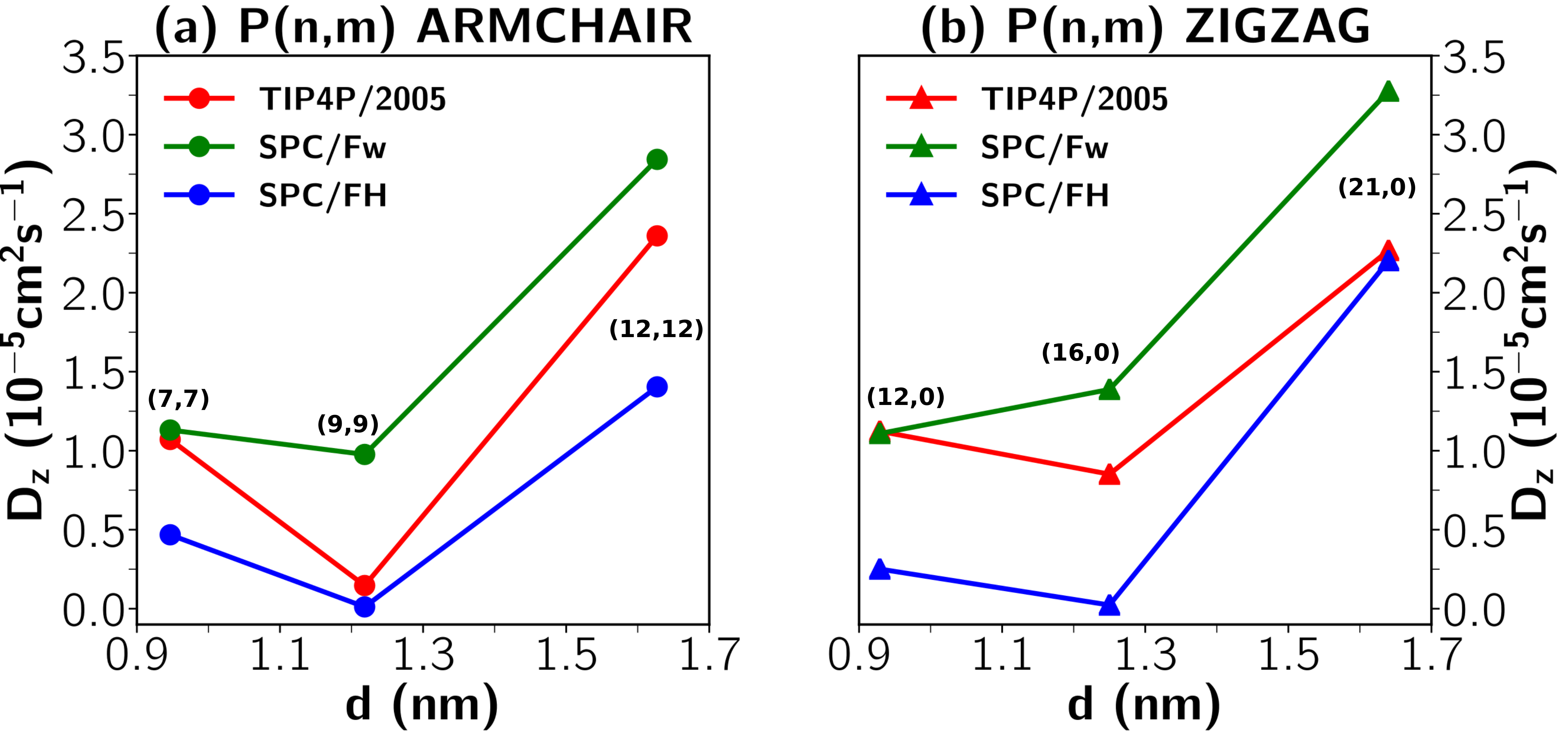}
	\end{center}
	\caption{Axial diffusion coefficient as a function of the diameter of perfect carbon nanotubes, P($n,m$) (a) armchair and (b) zigzag for T=$300$~K.} 
	\label{fig_diff-az}
\end{figure}

Then, we analyzed how the diffusion coefficient depends on the chirality by computing it for  P($n,m$) armchair and zigzag nanotubes for different diameters and water models. Figure~\ref{fig_diff-az} shows that water mobility is not strongly affected by CNT chirality for different diameters, except CNTs P($9,9$) and P($16,0$). These diameters are distinct when compared to smaller and larger nanotubes, because only a layer of water close to the wall is formed. This water layer presents a different water-wall interaction depending on the chirality~\cite{@10.1063/1.5129394,@10.1063/5.0031084,@10.1016/j.physa.2018.11.042}. Water is frozen within the CNT P($9,9$) for the TIP4P/2005 and SPC/FH models, while a non-zero diffusion is observed for the SPC/Fw model. In the case of CNT P($16,0$) a larger mobility was found for the SPC/Fw followed by the TIP4P/2005, while a low diffusion was observed for the SPC/FH case, close to that observed in the CNT P($9,9$). 

For the case of P($n,m$) nanotubes, armchairs and zigzags, the TIP4P/2005 and SPC/Fw water mobility are higher than SPC/FH what is expected since in this model water is more structured. In the cases of TIP4P/2005 and SPC/Fw water is frozen in the CNT P($9,9$), but not in CNT P($16,0$), despite both having the same diameter. The water melt in these two water models can be attributed to the spiral and ring-shaped structure water forms inside the armchair and zigzag tubes ~\cite{@10.1063/1.5129394}. SPC/FH water remains frozen for both CNT P($9,9$) and P($16,0$) what suggests that the more structured water might be less affected by the water-wall interaction.

Next, we addressed the question of how the deformations of CNTs affects the diffusion coefficient of water. In particular, we analyzed the difference in water diffusion for the armchair and zigzag CNTs when they change the structure from perfect P($n,m$) to wrinkled W($n,m$) and kneaded K($n,m$). We focused on two diameters of CNTs for each chirality, ($9,9$) and ($12,12$), for armchairs, and ($16,0$) and ($21,0$), for zigzags. The diameter selection was made to test how compression affects the number of layers and the distinct mobility between the two chirality depending on the chosen force field.

\begin{figure}[H]
  \begin{center}
  \includegraphics[width=6.4in]{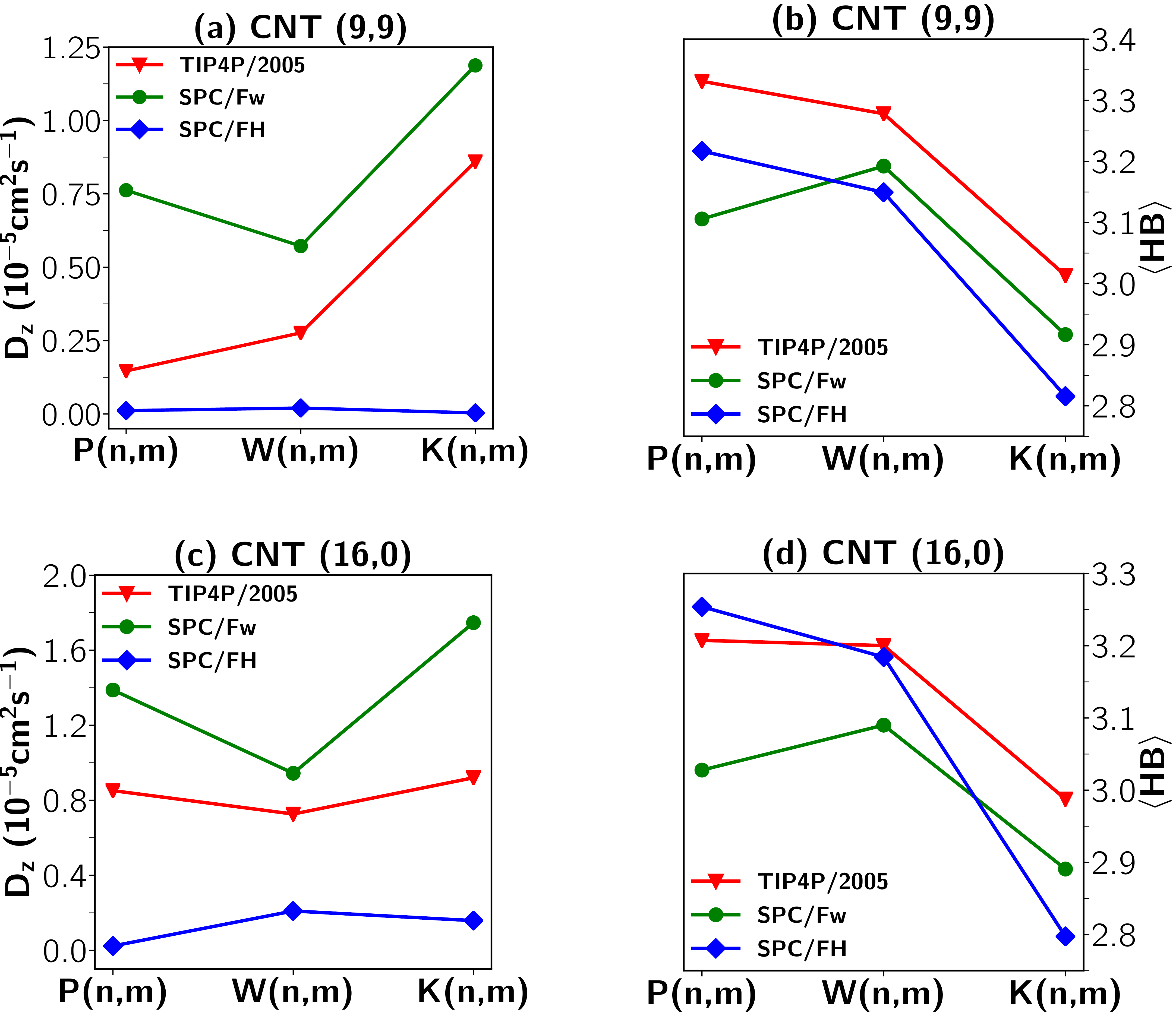}
  \end{center}
  \caption{Axial diffusion coefficient [left panels] and the average number of hydrogen bonds per water molecule [right panels] as a function of the kind of deformation of carbon nanotubes for (a), (b) ($9,9$) and (c), (d) ($16,0$), perfect P($n,m$), wrinkled W($n,m$) and kneaded K($n,m$) for water models TIP4P/2005, SPC/Fw and SPC/FH at T=$300$~K.}     
  \label{fig_diff-9-16}
\end{figure}

Figure~\ref{fig_diff-9-16} (a) shows that water mobility in the CNT ($9,9$) increases with both kind of applied strain only for the TIP4P/2005 model. For the SPC/Fw model, the mobility depends on the kind of applied strain, decreasing for wrinkled and increasing for kneaded. For the SPC/FH model, the water mobility is very small in comparison to those of the other models regardless of the strain. The behavior of the diffusion  is consistent with the number of hydrogen bonds formed. Figure~\ref{fig_diff-9-16} (b) for the  TIP4P/2005 and SPC/Fw where the increase (decrease) of the mobility is related to the decrease (increase) of the number of hydrogen bonds as it would be expected. For the SPC/FH, however, the diffusion is almost zero and the number of the hydrogen bonds decreases with the strain. Even thought distorting the nanotube disrupts the hydrogen bonds, this is not enough to melt the immobile water, which seems to find distinct ``ice-like'' structures as the system is compressed.

For the CNT ($16,0$), water diffusion, illustrated in Figure~\ref{fig_diff-9-16} (c), is less affected by the change from CNT by distortion for the TIP4P/2005 and SPC/FH models while the SPC/Fw presents the same increase followed by the decrease observed in the CNT ($9,9$). The decrease (increase) of the number of hydrogen bonds, shown in Figure~\ref{fig_diff-9-16} (d), is consistent with the increase (decrease) of the mobility for   TIP4P/2005 and SPC/Fw but not for the SPC/FH.  

The behavior of the diffusion coefficients of the water confined in the nanotubes P($n,m$) and W($n,m$) of the CNTs ($9,9$) and ($16,0$) are quite different for TIP4P/2005 and SPC/Fw models, indicating for this small diameter that the surface effects are really relevant. However, for the SPC/FH model, the water forms a frozen-like structure which does not depend strongly on the chirality as it will be shown by the density maps below.

\begin{figure}[H]
	\begin{center}
		\includegraphics[width=5.2in]{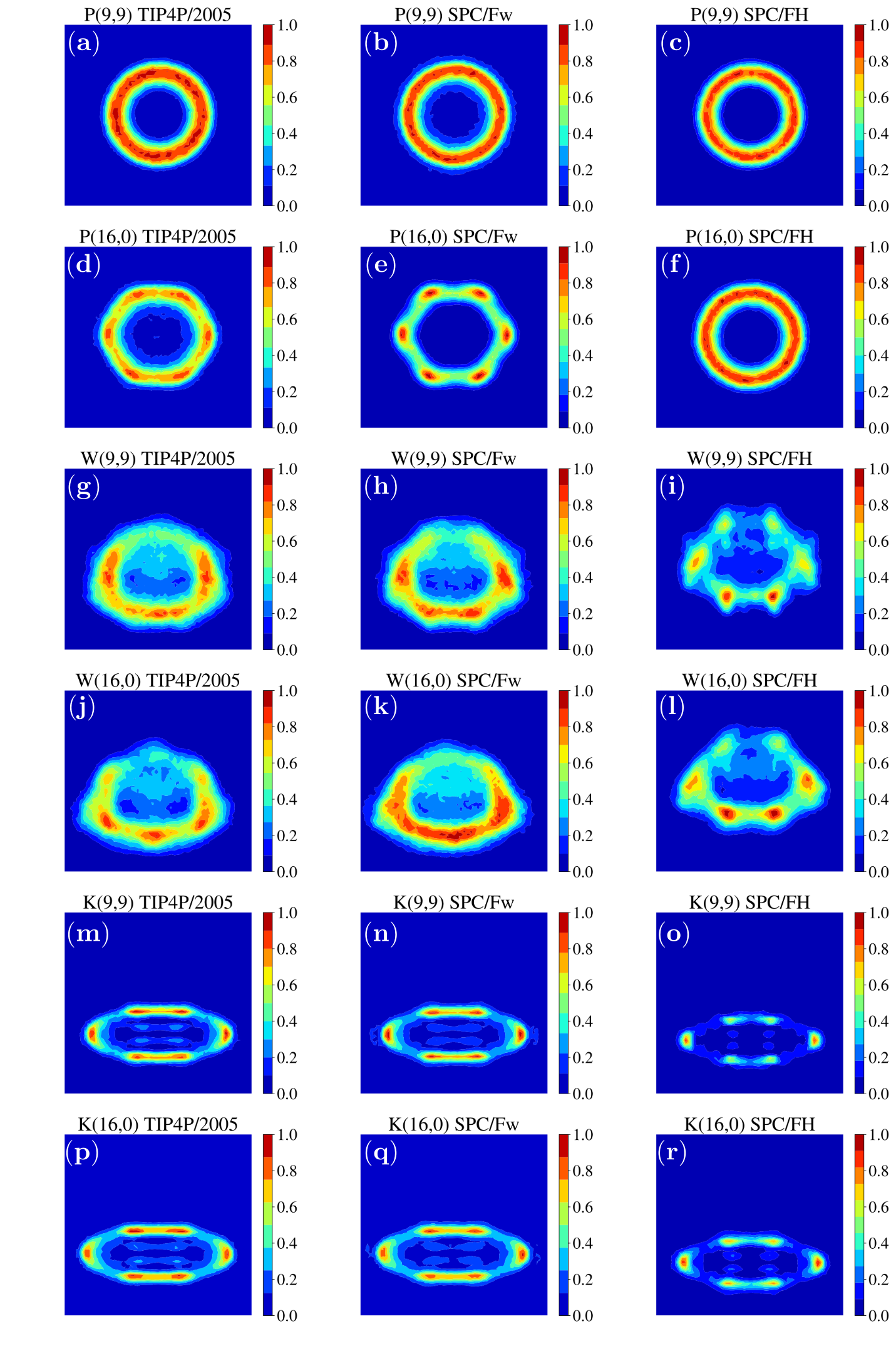}
	\end{center}
	\caption{Density maps in the $xy$ direction for the carbon nanotubes ($9,9$) and ($16,0$), perfect P($n,m$), wrinkled W($n,m$) and kneaded K($n,m$), and the comparison of the TIP4P/2005, SPC/Fw and SPC/FH water models. Dark blue regions have a low probability of finding water molecules, while red regions have a high probability of finding water molecules.}     
	\label{fig_maps-9x9-16x0-P}
\end{figure}

Finally, in order to understand the differences and similarities in CNT with distinct chirality and distortions, we analyzed the density  maps shown in Figure~\ref{fig_maps-9x9-16x0-P}. For the perfect CNT P($9,9$) in all water models analyzed, the water molecules are uniformly distributed in the vicinity of the nanotube wall, forming a ``frozen-like'' structure. This  led us to conclude that the flexibility of the models was not able to thaw the water molecules, and that the effect of the tube structure on the water molecules prevails. The variation of the diffusion coefficient, therefore, is only due to the effect of the degrees of freedom of the water molecule. This phenomenon is repeated for perfect CNT P($16,0$) for TIP4P/2005 and SPC/FW models, where water molecules assume a hexagonal distribution following the  wall boundary interaction. 

Water rearranges itself in the two types of CNTs is due to differences in wall structures combined with the hydrophobic nature of the carbon-water interaction, where water molecules form hydrogen bonds avoiding proximity to the wall of CNTs in a region where water interaction between carbon atoms is strongly repulsive~\cite{@10.1063/1.5129394}. In this case, different from the CNT P($9,9$) for TIP4P/2005 and SPC/Fw, the variation of the diffusion coefficient is not only due to the effect of the degrees of freedom of the water molecules, but also due to the effect of the chirality of the CNT ``breaking'' the structuring of water molecules, that is, being a barrier in the process of breaking and forming hydrogen bonds, acting directly in the increase of diffusion.

For the SPC/FH model at the perfect CNT ($9,9$) and ($16,0$) illustrated in Figure~\ref{fig_maps-9x9-16x0-P} water forms a helical structure regardless of the wall. The model adapt to form the arrangement which gives the more immobile structure. Figure~\ref{fig_maps-9x9-16x0-P} also shows that the strain leads to a disorganization of the helical structure in the ($9,9$) and of the hexagonal structure in the ($16,0$) case for the TIP4P/2005 and SPC/Fw models, resulting in different structure. In the case of the SPC/FH model, the distortion melts the helical structure present in both cases, generating organized lines which are independent of the chirality.

\begin{figure}[H]
  \begin{center}
  \includegraphics[width=6.4in]{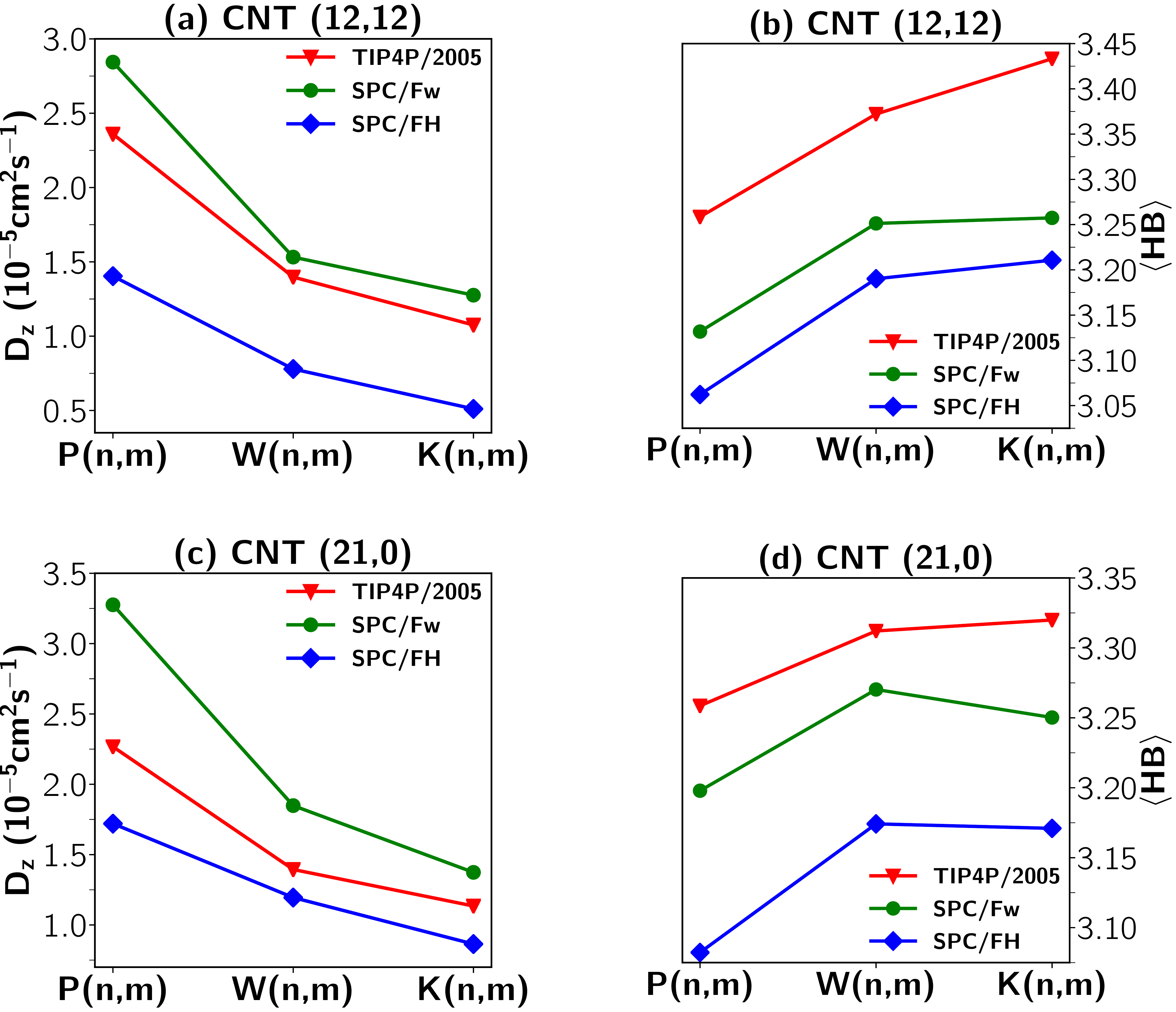}
  \end{center}
  \caption{Axial diffusion coefficient [left panels] and the average number of hydrogen bonds per water molecule [right panels] as a function of the kind of deformation of carbon nanotubes, ($12,12$) and ($21,0$), perfect P($n,m$), wrinkled W($n,m$) and kneaded K($n,m$) for water models TIP4P/2005, SPC/Fw and SPC/FH.}     
  \label{fig_diff-12-21}
\end{figure}

On last question is if the chirality independence observed in the SPC/FH model is present for larger diameters.
Figure~\ref{fig_diff-12-21} shows the diffusion coefficient versus the deformation for all the three models and the two chirality. 

Figures~\ref{fig_maps-9x9-16x0-P} (a) and (c) indicate that with the deformation of the CNT ($12,12$) there was a decrease in axial diffusion and there was an increase in the number of hydrogen bonds. The same behavior is observed in Figure~\ref{fig_maps-9x9-16x0-P} (b) and (d) for the CNT ($21,0$). 

In general, non-uniform deformations, as in W($n,m$) CNTs, bring water molecules closer together, favoring the formation of hydrogen bonds. This effect is even greater in K($n,m$) CNTs, in which the deformation and the decrease in the distance between the molecules are more uniform. Changes in water diffusion due to deformations in nanotubes ($12,12$) and ($21,0$) are very similar, and chirality seems to play a minor role in this case. For these CNTs, the water model, both the TIP4P/2005, the SCP/Fw and the SPC/FH, did not show relevant differences in mobility, because as the CNTs were deformed, the same behavior prevailed, decreasing the water diffusivity. Therefore, the diffusive behavior of water is more strongly affected due to structural deformations, making the degrees of freedom of water molecules play a secondary role.

\section{Conclusions}
\label{sec:conclusions}

In this work, we analyzed the diffusion coefficient of water under confinement in carbon nanotubes. Different nanotube sizes, topology and deformation were considered, as well as different theoretical models of water.

We observed that the choice of force field and the structure of CNTs directly affect the dynamic behavior of water. In addition, the diffusion coefficient for water confined in CNTs with different degrees of deformation showed a non-trivial behavior, with this we verified that the water will be dependent on the sum of the variation of the confining structure, the hydrophobic nature of the carbon nanotubes and the degrees of freedom of water molecules imposed by the force fields. 

The SPC/FH is a model that favors the water structure and suffers less impact from the wall structure for CNTs with smaller diameters. For larger CNTs, the influence of the walls becomes less relevant. In general, deformation suppresses mobility for large-sized CNTs and favors it for small-sized ones. These processes govern the mobility of confined water in a way that highlights the importance of choosing the force field of water in determining the transport properties of water.

\begin{acknowledgement}
This work is funded by the Brazilian scientific agency Conselho Nacional de Desenvolvimento Científico e Tecnológico (CNPq) and the Brazilian Institute of Science and Technology (INCT) in Carbon Nanomaterials with collaboration and computational support from Universidade Federal de Minas Gerais (UFMG), and Universidade Federal de Ouro Preto (UFOP). BHSM is grateful to Prof. Matheus J. S. Matos (UFOP) for discussions and theoretical contributions. ABO and RJCB thank the science agency FAPEMIG and PROPPI-UFOP for financial support. Authors acknowledge the National Laboratory for Scientific Computing (LNCC/MCTI, Brazil) for providing HPC resources of the SDumont supercomputer, which have contributed to the research results reported within this paper. URL: http://sdumont.lncc.br.
\end{acknowledgement}

\bibliography{achemso-demo}

\providecommand{\latin}[1]{#1}
\makeatletter
\providecommand{\doi}
  {\begingroup\let\do\@makeother\dospecials
  \catcode`\{=1 \catcode`\}=2 \doi@aux}
\providecommand{\doi@aux}[1]{\endgroup\texttt{#1}}
\makeatother
\providecommand*\mcitethebibliography{\thebibliography}
\csname @ifundefined\endcsname{endmcitethebibliography}
  {\let\endmcitethebibliography\endthebibliography}{}
\begin{mcitethebibliography}{68}
\providecommand*\natexlab[1]{#1}
\providecommand*\mciteSetBstSublistMode[1]{}
\providecommand*\mciteSetBstMaxWidthForm[2]{}
\providecommand*\mciteBstWouldAddEndPuncttrue
  {\def\EndOfBibitem{\unskip.}}
\providecommand*\mciteBstWouldAddEndPunctfalse
  {\let\EndOfBibitem\relax}
\providecommand*\mciteSetBstMidEndSepPunct[3]{}
\providecommand*\mciteSetBstSublistLabelBeginEnd[3]{}
\providecommand*\EndOfBibitem{}
\mciteSetBstSublistMode{f}
\mciteSetBstMaxWidthForm{subitem}{(\alph{mcitesubitemcount})}
\mciteSetBstSublistLabelBeginEnd
  {\mcitemaxwidthsubitemform\space}
  {\relax}
  {\relax}

\bibitem[Hummer \latin{et~al.}(2001)Hummer, Rasaiah, and
  Noworyta]{@nature/35102535}
Hummer,~G.; Rasaiah,~J.~C.; Noworyta,~J.~P. Water conduction through the
  hydrophobic channel of a carbon nanotube. \emph{Nature} \textbf{2001},
  \emph{414}, 188--190\relax
\mciteBstWouldAddEndPuncttrue
\mciteSetBstMidEndSepPunct{\mcitedefaultmidpunct}
{\mcitedefaultendpunct}{\mcitedefaultseppunct}\relax
\EndOfBibitem
\bibitem[Holt \latin{et~al.}(2004)Holt, Noy, Huser, Eaglesham, and
  Bakajin]{@10.1021/nl048876h}
Holt,~J.~K.; Noy,~A.; Huser,~T.; Eaglesham,~D.; Bakajin,~O. Fabrication of a
  carbon nanotube-embedded silicon nitride membrane for studies of
  nanometer-scale mass transport. \emph{Nano Letters} \textbf{2004}, \emph{4},
  2245--2250\relax
\mciteBstWouldAddEndPuncttrue
\mciteSetBstMidEndSepPunct{\mcitedefaultmidpunct}
{\mcitedefaultendpunct}{\mcitedefaultseppunct}\relax
\EndOfBibitem
\bibitem[Majumder \latin{et~al.}(2005)Majumder, Chopra, Andrews, and
  Hinds]{@10.1038/43844a}
Majumder,~M.; Chopra,~N.; Andrews,~R.; Hinds,~B. Erratum: Nanoscale
  hydrodynamics: Enhanced flow in carbon nanotubes. \emph{Nature}
  \textbf{2005}, \emph{438}, 930--930\relax
\mciteBstWouldAddEndPuncttrue
\mciteSetBstMidEndSepPunct{\mcitedefaultmidpunct}
{\mcitedefaultendpunct}{\mcitedefaultseppunct}\relax
\EndOfBibitem
\bibitem[Striolo(2006)]{@10.1021/nl052254u}
Striolo,~A. The mechanism of water diffusion in narrow carbon nanotubes.
  \emph{Nano Letters} \textbf{2006}, \emph{6}, 633--639\relax
\mciteBstWouldAddEndPuncttrue
\mciteSetBstMidEndSepPunct{\mcitedefaultmidpunct}
{\mcitedefaultendpunct}{\mcitedefaultseppunct}\relax
\EndOfBibitem
\bibitem[Whitby \latin{et~al.}(2008)Whitby, Cagnon, Thanou, and
  Quirke]{@10.1021/nl080705f}
Whitby,~M.; Cagnon,~L.; Thanou,~M.; Quirke,~N. Enhanced fluid flow through
  nanoscale carbon pipes. \emph{Nano Letters} \textbf{2008}, \emph{8},
  2632--2637\relax
\mciteBstWouldAddEndPuncttrue
\mciteSetBstMidEndSepPunct{\mcitedefaultmidpunct}
{\mcitedefaultendpunct}{\mcitedefaultseppunct}\relax
\EndOfBibitem
\bibitem[Zhang \latin{et~al.}(2011)Zhang, Ye, Zheng, and
  Zhang]{@10.1007/s10404-010-0678-0}
Zhang,~H.; Ye,~H.; Zheng,~Y.; Zhang,~Z. Prediction of the viscosity of water
  confined in carbon nanotubes. \emph{Microfluidics and Nanofluidics}
  \textbf{2011}, \emph{10}, 403--414\relax
\mciteBstWouldAddEndPuncttrue
\mciteSetBstMidEndSepPunct{\mcitedefaultmidpunct}
{\mcitedefaultendpunct}{\mcitedefaultseppunct}\relax
\EndOfBibitem
\bibitem[Secchi \latin{et~al.}(2016)Secchi, Marbach, Nigu{\`e}s, Stein, Siria,
  and Bocquet]{@10.1038/nature19315}
Secchi,~E.; Marbach,~S.; Nigu{\`e}s,~A.; Stein,~D.; Siria,~A.; Bocquet,~L.
  Massive radius-dependent flow slippage in carbon nanotubes. \emph{Nature}
  \textbf{2016}, \emph{537}, 210--213\relax
\mciteBstWouldAddEndPuncttrue
\mciteSetBstMidEndSepPunct{\mcitedefaultmidpunct}
{\mcitedefaultendpunct}{\mcitedefaultseppunct}\relax
\EndOfBibitem
\bibitem[Kashyap \latin{et~al.}(2020)Kashyap, Yang, and
  Datta]{@10.48550/arXiv.2001.00614}
Kashyap,~J.; Yang,~E.-H.; Datta,~D. Comprehensive understanding of water-driven
  graphene wrinkle life-cycle towards applications in flexible electronics: A
  computational study. \emph{ArXiv Preprint ArXiv:2001.00614} \textbf{2020},
  23\relax
\mciteBstWouldAddEndPuncttrue
\mciteSetBstMidEndSepPunct{\mcitedefaultmidpunct}
{\mcitedefaultendpunct}{\mcitedefaultseppunct}\relax
\EndOfBibitem
\bibitem[Mendon{\c{c}}a \latin{et~al.}(2019)Mendon{\c{c}}a, de~Freitas,
  K{\"o}hler, Batista, Barbosa, and de~Oliveira]{@10.1016/j.physa.2018.11.042}
Mendon{\c{c}}a,~B. H.~S.; de~Freitas,~D.~N.; K{\"o}hler,~M.~H.; Batista,~R.~J.;
  Barbosa,~M.~C.; de~Oliveira,~A.~B. Diffusion behaviour of water confined in
  deformed carbon nanotubes. \emph{Physica A: Statistical Mechanics and its
  Applications} \textbf{2019}, \emph{517}, 491--498\relax
\mciteBstWouldAddEndPuncttrue
\mciteSetBstMidEndSepPunct{\mcitedefaultmidpunct}
{\mcitedefaultendpunct}{\mcitedefaultseppunct}\relax
\EndOfBibitem
\bibitem[Mendon{\c{c}}a \latin{et~al.}(2020)Mendon{\c{c}}a, Ternes, Salcedo,
  de~Oliveira, and Barbosa]{@10.1063/1.5129394}
Mendon{\c{c}}a,~B. H.~S.; Ternes,~P.; Salcedo,~E.; de~Oliveira,~A.~B.;
  Barbosa,~M.~C. Water diffusion in rough carbon nanotubes. \emph{The Journal
  of Chemical Physics} \textbf{2020}, \emph{152}, 024708\relax
\mciteBstWouldAddEndPuncttrue
\mciteSetBstMidEndSepPunct{\mcitedefaultmidpunct}
{\mcitedefaultendpunct}{\mcitedefaultseppunct}\relax
\EndOfBibitem
\bibitem[de~Freitas \latin{et~al.}(2020)de~Freitas, Mendon{\c{c}}a, K{\"o}hler,
  Barbosa, Matos, Batista, and de~Oliveira]{@10.1016/j.chemphys.2020.110849}
de~Freitas,~D.~N.; Mendon{\c{c}}a,~B. H.~S.; K{\"o}hler,~M.~H.; Barbosa,~M.~C.;
  Matos,~M.~J.; Batista,~R.~J.; de~Oliveira,~A.~B. Water diffusion in carbon
  nanotubes under directional electric frields: Coupling between mobility and
  hydrogen bonding. \emph{Chemical Physics} \textbf{2020}, \emph{537},
  110849\relax
\mciteBstWouldAddEndPuncttrue
\mciteSetBstMidEndSepPunct{\mcitedefaultmidpunct}
{\mcitedefaultendpunct}{\mcitedefaultseppunct}\relax
\EndOfBibitem
\bibitem[Thiemann \latin{et~al.}(2022)Thiemann, Schran, Rowe, M{\"u}ller, and
  Michaelides]{@10.1021/acsnano.2c02784}
Thiemann,~F.~L.; Schran,~C.; Rowe,~P.; M{\"u}ller,~E.~A.; Michaelides,~A. Water
  flow in single-wall nanotubes: Oxygen makes it slip, hydrogen makes it stick.
  \emph{ACS Nano} \textbf{2022}, \relax
\mciteBstWouldAddEndPunctfalse
\mciteSetBstMidEndSepPunct{\mcitedefaultmidpunct}
{}{\mcitedefaultseppunct}\relax
\EndOfBibitem
\bibitem[d’Eurydice and Galvosas(2014)d’Eurydice, and
  Galvosas]{@10.1016/j.jmr.2014.07.012}
d’Eurydice,~M.~N.; Galvosas,~P. Measuring diffusion--relaxation correlation
  maps using non-uniform field gradients of single-sided NMR devices.
  \emph{Journal of Magnetic Resonance} \textbf{2014}, \emph{248},
  137--145\relax
\mciteBstWouldAddEndPuncttrue
\mciteSetBstMidEndSepPunct{\mcitedefaultmidpunct}
{\mcitedefaultendpunct}{\mcitedefaultseppunct}\relax
\EndOfBibitem
\bibitem[Liu \latin{et~al.}(2010)Liu, He, Tang, Liu, Pang, Cao, Krstic, Joseph,
  Lindsay, and Nuckolls]{@10.1126/science.1181799}
Liu,~H.; He,~J.; Tang,~J.; Liu,~H.; Pang,~P.; Cao,~D.; Krstic,~P.; Joseph,~S.;
  Lindsay,~S.; Nuckolls,~C. Translocation of single-stranded DNA through
  single-walled carbon nanotubes. \emph{Science} \textbf{2010}, \emph{327},
  64--67\relax
\mciteBstWouldAddEndPuncttrue
\mciteSetBstMidEndSepPunct{\mcitedefaultmidpunct}
{\mcitedefaultendpunct}{\mcitedefaultseppunct}\relax
\EndOfBibitem
\bibitem[Bocquet and Charlaix(2010)Bocquet, and Charlaix]{@10.1039/B909366B}
Bocquet,~L.; Charlaix,~E. Nanofluidics, from bulk to interfaces. \emph{Chemical
  Society Reviews} \textbf{2010}, \emph{39}, 1073--1095\relax
\mciteBstWouldAddEndPuncttrue
\mciteSetBstMidEndSepPunct{\mcitedefaultmidpunct}
{\mcitedefaultendpunct}{\mcitedefaultseppunct}\relax
\EndOfBibitem
\bibitem[Wu \latin{et~al.}(2010)Wu, Paudel, Strasinger, Hammell, Stinchcomb,
  and Hinds]{@10.1073/pnas.1004714107}
Wu,~J.; Paudel,~K.~S.; Strasinger,~C.; Hammell,~D.; Stinchcomb,~A.~L.;
  Hinds,~B.~J. Programmable transdermal drug delivery of nicotine using carbon
  nanotube membranes. \emph{Proceedings of the National Academy of Sciences}
  \textbf{2010}, \emph{107}, 11698--11702\relax
\mciteBstWouldAddEndPuncttrue
\mciteSetBstMidEndSepPunct{\mcitedefaultmidpunct}
{\mcitedefaultendpunct}{\mcitedefaultseppunct}\relax
\EndOfBibitem
\bibitem[Elimelech and Phillip(2011)Elimelech, and
  Phillip]{@10.1126/science.1200488}
Elimelech,~M.; Phillip,~W.~A. The future of seawater desalination: energy,
  technology, and the environment. \emph{Science} \textbf{2011}, \emph{333},
  712--717\relax
\mciteBstWouldAddEndPuncttrue
\mciteSetBstMidEndSepPunct{\mcitedefaultmidpunct}
{\mcitedefaultendpunct}{\mcitedefaultseppunct}\relax
\EndOfBibitem
\bibitem[Zhang \latin{et~al.}(2011)Zhang, Zhang, and
  Zhang]{@10.1186/1556-276X-6-555}
Zhang,~W.; Zhang,~Z.; Zhang,~Y. The application of carbon nanotubes in target
  drug delivery systems for cancer therapies. \emph{Nanoscale Research Letters}
  \textbf{2011}, \emph{6}, 1--22\relax
\mciteBstWouldAddEndPuncttrue
\mciteSetBstMidEndSepPunct{\mcitedefaultmidpunct}
{\mcitedefaultendpunct}{\mcitedefaultseppunct}\relax
\EndOfBibitem
\bibitem[Logan and Elimelech(2012)Logan, and Elimelech]{@10.1038/nature11477}
Logan,~B.~E.; Elimelech,~M. Membrane-based processes for sustainable power
  generation using water. \emph{Nature} \textbf{2012}, \emph{488},
  313--319\relax
\mciteBstWouldAddEndPuncttrue
\mciteSetBstMidEndSepPunct{\mcitedefaultmidpunct}
{\mcitedefaultendpunct}{\mcitedefaultseppunct}\relax
\EndOfBibitem
\bibitem[Surwade \latin{et~al.}(2015)Surwade, Smirnov, Vlassiouk, Unocic,
  Veith, Dai, and Mahurin]{@10.1038/NNANO.2015.37}
Surwade,~S.~P.; Smirnov,~S.~N.; Vlassiouk,~I.~V.; Unocic,~R.~R.; Veith,~G.~M.;
  Dai,~S.; Mahurin,~S.~M. Water desalination using nanoporous single-layer
  graphene. \emph{Nature Nanotechnology} \textbf{2015}, \emph{10},
  459--464\relax
\mciteBstWouldAddEndPuncttrue
\mciteSetBstMidEndSepPunct{\mcitedefaultmidpunct}
{\mcitedefaultendpunct}{\mcitedefaultseppunct}\relax
\EndOfBibitem
\bibitem[Park and Jung(2014)Park, and Jung]{@10.1039/C3CS60253B}
Park,~H.~G.; Jung,~Y. Carbon nanofluidics of rapid water transport for energy
  applications. \emph{Chemical Society Reviews} \textbf{2014}, \emph{43},
  565--576\relax
\mciteBstWouldAddEndPuncttrue
\mciteSetBstMidEndSepPunct{\mcitedefaultmidpunct}
{\mcitedefaultendpunct}{\mcitedefaultseppunct}\relax
\EndOfBibitem
\bibitem[Ketabi and Rahmani(2017)Ketabi, and
  Rahmani]{@10.1016/j.msec.2016.12.058}
Ketabi,~S.; Rahmani,~L. Carbon nanotube as a carrier in drug delivery system
  for carnosine dipeptide: A computer simulation study. \emph{Materials Science
  and Engineering: C} \textbf{2017}, \emph{73}, 173--181\relax
\mciteBstWouldAddEndPuncttrue
\mciteSetBstMidEndSepPunct{\mcitedefaultmidpunct}
{\mcitedefaultendpunct}{\mcitedefaultseppunct}\relax
\EndOfBibitem
\bibitem[Bocquet(2020)]{@10.1038/s41563-020-0625-8}
Bocquet,~L. Nanofluidics coming of age. \emph{Nature Materials} \textbf{2020},
  \emph{19}, 254--256\relax
\mciteBstWouldAddEndPuncttrue
\mciteSetBstMidEndSepPunct{\mcitedefaultmidpunct}
{\mcitedefaultendpunct}{\mcitedefaultseppunct}\relax
\EndOfBibitem
\bibitem[Maller \latin{et~al.}(2021)Maller, Drain, Barrett, Borgquist, Ruffell,
  Zakharevich, Pham, Gruosso, Kuasne, Lakins, \latin{et~al.}
  others]{@10.1038/s41563-020-00849-5}
Maller,~O.; Drain,~A.~P.; Barrett,~A.~S.; Borgquist,~S.; Ruffell,~B.;
  Zakharevich,~I.; Pham,~T.~T.; Gruosso,~T.; Kuasne,~H.; Lakins,~J.~N.,
  \latin{et~al.}  Tumour-associated macrophages drive stromal cell-dependent
  collagen crosslinking and stiffening to promote breast cancer aggression.
  \emph{Nature Materials} \textbf{2021}, \emph{20}, 548--559\relax
\mciteBstWouldAddEndPuncttrue
\mciteSetBstMidEndSepPunct{\mcitedefaultmidpunct}
{\mcitedefaultendpunct}{\mcitedefaultseppunct}\relax
\EndOfBibitem
\bibitem[Striolo(2007)]{@10.1088/0957-4484/18/47/475704}
Striolo,~A. Water self-diffusion through narrow oxygenated carbon nanotubes.
  \emph{Nanotechnology} \textbf{2007}, \emph{18}, 475704\relax
\mciteBstWouldAddEndPuncttrue
\mciteSetBstMidEndSepPunct{\mcitedefaultmidpunct}
{\mcitedefaultendpunct}{\mcitedefaultseppunct}\relax
\EndOfBibitem
\bibitem[Pascal \latin{et~al.}(2011)Pascal, Goddard, and
  Jung]{@10.1073/pnas.1108073108}
Pascal,~T.~A.; Goddard,~W.~A.; Jung,~Y. Entropy and the driving force for the
  filling of carbon nanotubes with water. \emph{Proceedings of the National
  Academy of Sciences} \textbf{2011}, \emph{108}, 11794--11798\relax
\mciteBstWouldAddEndPuncttrue
\mciteSetBstMidEndSepPunct{\mcitedefaultmidpunct}
{\mcitedefaultendpunct}{\mcitedefaultseppunct}\relax
\EndOfBibitem
\bibitem[Wei \latin{et~al.}(2000)Wei, Bechinger, and
  Leiderer]{@10.1126/science.287.5453.625}
Wei,~Q.-H.; Bechinger,~C.; Leiderer,~P. Single-file diffusion of colloids in
  one-dimensional channels. \emph{Science} \textbf{2000}, \emph{287},
  625--627\relax
\mciteBstWouldAddEndPuncttrue
\mciteSetBstMidEndSepPunct{\mcitedefaultmidpunct}
{\mcitedefaultendpunct}{\mcitedefaultseppunct}\relax
\EndOfBibitem
\bibitem[Liu \latin{et~al.}(2005)Liu, Wang, Wu, and Zhang]{@10.1063/1.2131070}
Liu,~Y.; Wang,~Q.; Wu,~T.; Zhang,~L. Fluid structure and transport properties
  of water inside carbon nanotubes. \emph{The Journal of Chemical Physics}
  \textbf{2005}, \emph{123}, 234701\relax
\mciteBstWouldAddEndPuncttrue
\mciteSetBstMidEndSepPunct{\mcitedefaultmidpunct}
{\mcitedefaultendpunct}{\mcitedefaultseppunct}\relax
\EndOfBibitem
\bibitem[Mukherjee \latin{et~al.}(2007)Mukherjee, Maiti, Dasgupta, and
  Sood]{@10.1063/1.2565806}
Mukherjee,~B.; Maiti,~P.~K.; Dasgupta,~C.; Sood,~A. Strong correlations and
  Fickian water diffusion in narrow carbon nanotubes. \emph{The Journal of
  Chemical Physics} \textbf{2007}, \emph{126}, 124704\relax
\mciteBstWouldAddEndPuncttrue
\mciteSetBstMidEndSepPunct{\mcitedefaultmidpunct}
{\mcitedefaultendpunct}{\mcitedefaultseppunct}\relax
\EndOfBibitem
\bibitem[Ye \latin{et~al.}(2011)Ye, Zhang, Zheng, and
  Zhang]{@10.1007/s10404-011-0772-y}
Ye,~H.; Zhang,~H.; Zheng,~Y.; Zhang,~Z. Nanoconfinement induced anomalous water
  diffusion inside carbon nanotubes. \emph{Microfluidics and Nanofluidics}
  \textbf{2011}, \emph{10}, 1359--1364\relax
\mciteBstWouldAddEndPuncttrue
\mciteSetBstMidEndSepPunct{\mcitedefaultmidpunct}
{\mcitedefaultendpunct}{\mcitedefaultseppunct}\relax
\EndOfBibitem
\bibitem[Barati~Farimani and Aluru(2011)Barati~Farimani, and
  Aluru]{@10.1021/jp205877b}
Barati~Farimani,~A.; Aluru,~N.~R. Spatial diffusion of water in carbon
  nanotubes: from fickian to ballistic motion. \emph{The Journal of Physical
  Chemistry B} \textbf{2011}, \emph{115}, 12145--12149\relax
\mciteBstWouldAddEndPuncttrue
\mciteSetBstMidEndSepPunct{\mcitedefaultmidpunct}
{\mcitedefaultendpunct}{\mcitedefaultseppunct}\relax
\EndOfBibitem
\bibitem[Berendsen \latin{et~al.}(1981)Berendsen, Postma, van Gunsteren, and
  Hermans]{@10.1007/978-94-015-7658-1_21}
Berendsen,~H.~J.; Postma,~J.~P.; van Gunsteren,~W.~F.; Hermans,~J.
  \emph{Intermolecular Forces}; 1981; pp 331--342\relax
\mciteBstWouldAddEndPuncttrue
\mciteSetBstMidEndSepPunct{\mcitedefaultmidpunct}
{\mcitedefaultendpunct}{\mcitedefaultseppunct}\relax
\EndOfBibitem
\bibitem[Kaminski \latin{et~al.}(2001)Kaminski, Friesner, Tirado-Rives, and
  Jorgensen]{@10.1021/jp003919d}
Kaminski,~G.~A.; Friesner,~R.~A.; Tirado-Rives,~J.; Jorgensen,~W.~L. Evaluation
  and reparametrization of the OPLS-AA force field for proteins via comparison
  with accurate quantum chemical calculations on peptides. \emph{The Journal of
  Physical Chemistry B} \textbf{2001}, \emph{105}, 6474--6487\relax
\mciteBstWouldAddEndPuncttrue
\mciteSetBstMidEndSepPunct{\mcitedefaultmidpunct}
{\mcitedefaultendpunct}{\mcitedefaultseppunct}\relax
\EndOfBibitem
\bibitem[Horn \latin{et~al.}(2004)Horn, Swope, Pitera, Madura, Dick, Hura, and
  Head-Gordon]{@10.1063/1.1683075}
Horn,~H.~W.; Swope,~W.~C.; Pitera,~J.~W.; Madura,~J.~D.; Dick,~T.~J.;
  Hura,~G.~L.; Head-Gordon,~T. Development of an improved four-site water model
  for biomolecular simulations: TIP4P-Ew. \emph{The Journal of Chemical
  Physics} \textbf{2004}, \emph{120}, 9665--9678\relax
\mciteBstWouldAddEndPuncttrue
\mciteSetBstMidEndSepPunct{\mcitedefaultmidpunct}
{\mcitedefaultendpunct}{\mcitedefaultseppunct}\relax
\EndOfBibitem
\bibitem[Jorgensen \latin{et~al.}(1983)Jorgensen, Chandrasekhar, Madura, Impey,
  and Klein]{@10.1063/1.445869}
Jorgensen,~W.~L.; Chandrasekhar,~J.; Madura,~J.~D.; Impey,~R.~W.; Klein,~M.~L.
  Comparison of simple potential functions for simulating liquid water.
  \emph{The Journal of Chemical Physics} \textbf{1983}, \emph{79},
  926--935\relax
\mciteBstWouldAddEndPuncttrue
\mciteSetBstMidEndSepPunct{\mcitedefaultmidpunct}
{\mcitedefaultendpunct}{\mcitedefaultseppunct}\relax
\EndOfBibitem
\bibitem[Mahoney and Jorgensen(2000)Mahoney, and Jorgensen]{@10.1063/1.481505}
Mahoney,~M.~W.; Jorgensen,~W.~L. A five-site model for liquid water and the
  reproduction of the density anomaly by rigid, nonpolarizable potential
  functions. \emph{The Journal of Chemical Physics} \textbf{2000}, \emph{112},
  8910--8922\relax
\mciteBstWouldAddEndPuncttrue
\mciteSetBstMidEndSepPunct{\mcitedefaultmidpunct}
{\mcitedefaultendpunct}{\mcitedefaultseppunct}\relax
\EndOfBibitem
\bibitem[Vega and Abascal(2005)Vega, and Abascal]{@10.1063/1.2056539}
Vega,~C.; Abascal,~J. Relation between the melting temperature and the
  temperature of maximum density for the most common models of water. \emph{The
  Journal of Chemical Physics} \textbf{2005}, \emph{123}, 144504\relax
\mciteBstWouldAddEndPuncttrue
\mciteSetBstMidEndSepPunct{\mcitedefaultmidpunct}
{\mcitedefaultendpunct}{\mcitedefaultseppunct}\relax
\EndOfBibitem
\bibitem[L{\'o}pez-Lemus \latin{et~al.}(2008)L{\'o}pez-Lemus, Chapela, and
  Alejandre]{@10.1063/1.2907845}
L{\'o}pez-Lemus,~J.; Chapela,~G.~A.; Alejandre,~J. Effect of flexibility on
  surface tension and coexisting densities of water. \emph{The Journal of
  Chemical Physics} \textbf{2008}, \emph{128}, 174703\relax
\mciteBstWouldAddEndPuncttrue
\mciteSetBstMidEndSepPunct{\mcitedefaultmidpunct}
{\mcitedefaultendpunct}{\mcitedefaultseppunct}\relax
\EndOfBibitem
\bibitem[Alejandre \latin{et~al.}(2009)Alejandre, Chapela, Bresme, and
  Hansen]{@10.1063/1.3124184}
Alejandre,~J.; Chapela,~G.~A.; Bresme,~F.; Hansen,~J.-P. The short range
  anion-H interaction is the driving force for crystal formation of ions in
  water. \emph{The Journal of Chemical Physics} \textbf{2009}, \emph{130},
  174505\relax
\mciteBstWouldAddEndPuncttrue
\mciteSetBstMidEndSepPunct{\mcitedefaultmidpunct}
{\mcitedefaultendpunct}{\mcitedefaultseppunct}\relax
\EndOfBibitem
\bibitem[Zhang \latin{et~al.}(2021)Zhang, Wang, Car, and
  Weinan]{@10.1103/PhysRevLett.126.236001}
Zhang,~L.; Wang,~H.; Car,~R.; Weinan,~E. Phase diagram of a deep potential
  water model. \emph{Physical Review Letters} \textbf{2021}, \emph{126},
  236001\relax
\mciteBstWouldAddEndPuncttrue
\mciteSetBstMidEndSepPunct{\mcitedefaultmidpunct}
{\mcitedefaultendpunct}{\mcitedefaultseppunct}\relax
\EndOfBibitem
\bibitem[Wallqvist and Teleman(1991)Wallqvist, and
  Teleman]{@10.1080/00268979100102391}
Wallqvist,~A.; Teleman,~O. Properties of flexible water models. \emph{Molecular
  Physics} \textbf{1991}, \emph{74}, 515--533\relax
\mciteBstWouldAddEndPuncttrue
\mciteSetBstMidEndSepPunct{\mcitedefaultmidpunct}
{\mcitedefaultendpunct}{\mcitedefaultseppunct}\relax
\EndOfBibitem
\bibitem[Tocci \latin{et~al.}(2020)Tocci, Bilichenko, Joly, and
  Iannuzzi]{@10.1039/D0NR02511A}
Tocci,~G.; Bilichenko,~M.; Joly,~L.; Iannuzzi,~M. Ab initio nanofluidics:
  disentangling the role of the energy landscape and of density correlations on
  liquid/solid friction. \emph{Nanoscale} \textbf{2020}, \emph{12},
  10994--11000\relax
\mciteBstWouldAddEndPuncttrue
\mciteSetBstMidEndSepPunct{\mcitedefaultmidpunct}
{\mcitedefaultendpunct}{\mcitedefaultseppunct}\relax
\EndOfBibitem
\bibitem[Abascal and Vega(2005)Abascal, and Vega]{@10.1063/1.2121687}
Abascal,~J.~L.; Vega,~C. A general purpose model for the condensed phases of
  water: TIP4P/2005. \emph{The Journal of Chemical Physics} \textbf{2005},
  \emph{123}, 234505\relax
\mciteBstWouldAddEndPuncttrue
\mciteSetBstMidEndSepPunct{\mcitedefaultmidpunct}
{\mcitedefaultendpunct}{\mcitedefaultseppunct}\relax
\EndOfBibitem
\bibitem[Teleman \latin{et~al.}(1987)Teleman, J{\"o}nsson, and
  Engstr{\"o}m]{@10.1080/00268978700100141}
Teleman,~O.; J{\"o}nsson,~B.; Engstr{\"o}m,~S. A molecular dynamics simulation
  of a water model with intramolecular degrees of freedom. \emph{Molecular
  Physics} \textbf{1987}, \emph{60}, 193--203\relax
\mciteBstWouldAddEndPuncttrue
\mciteSetBstMidEndSepPunct{\mcitedefaultmidpunct}
{\mcitedefaultendpunct}{\mcitedefaultseppunct}\relax
\EndOfBibitem
\bibitem[Wu \latin{et~al.}(2006)Wu, Tepper, and Voth]{@10.1063/1.2136877}
Wu,~Y.; Tepper,~H.~L.; Voth,~G.~A. Flexible simple point-charge water model
  with improved liquid-state properties. \emph{The Journal of Chemical Physics}
  \textbf{2006}, \emph{124}, 024503\relax
\mciteBstWouldAddEndPuncttrue
\mciteSetBstMidEndSepPunct{\mcitedefaultmidpunct}
{\mcitedefaultendpunct}{\mcitedefaultseppunct}\relax
\EndOfBibitem
\bibitem[Chen \latin{et~al.}(1998)Chen, Hamon, Hu, Chen, Rao, Eklund, and
  Haddon]{@10.1126/science.282.5386.95}
Chen,~J.; Hamon,~M.~A.; Hu,~H.; Chen,~Y.; Rao,~A.~M.; Eklund,~P.~C.;
  Haddon,~R.~C. Solution properties of single-walled carbon nanotubes.
  \emph{Science} \textbf{1998}, \emph{282}, 95--98\relax
\mciteBstWouldAddEndPuncttrue
\mciteSetBstMidEndSepPunct{\mcitedefaultmidpunct}
{\mcitedefaultendpunct}{\mcitedefaultseppunct}\relax
\EndOfBibitem
\bibitem[Dalla~Bernardina \latin{et~al.}(2016)Dalla~Bernardina, Paineau,
  Brubach, Judeinstein, Rouzi{\`e}re, Launois, and Roy]{@10.1021/jacs.6b02635}
Dalla~Bernardina,~S.; Paineau,~E.; Brubach,~J.-B.; Judeinstein,~P.;
  Rouzi{\`e}re,~S.; Launois,~P.; Roy,~P. Water in carbon nanotubes: the
  peculiar hydrogen bond network revealed by infrared spectroscopy.
  \emph{Journal of the American Chemical Society} \textbf{2016}, \emph{138},
  10437--10443\relax
\mciteBstWouldAddEndPuncttrue
\mciteSetBstMidEndSepPunct{\mcitedefaultmidpunct}
{\mcitedefaultendpunct}{\mcitedefaultseppunct}\relax
\EndOfBibitem
\bibitem[Kukovecz \latin{et~al.}(2002)Kukovecz, Kramberger, Georgakilas, Prato,
  and Kuzmany]{@10.1140/epjb/e2002-00224-8}
Kukovecz,~A.; Kramberger,~C.; Georgakilas,~V.; Prato,~M.; Kuzmany,~H. A
  detailed Raman study on thin single-wall carbon nanotubes prepared by the
  HiPCO process. \emph{The European Physical Journal B-Condensed Matter and
  Complex Systems} \textbf{2002}, \emph{28}, 223--230\relax
\mciteBstWouldAddEndPuncttrue
\mciteSetBstMidEndSepPunct{\mcitedefaultmidpunct}
{\mcitedefaultendpunct}{\mcitedefaultseppunct}\relax
\EndOfBibitem
\bibitem[Mamontov \latin{et~al.}(2006)Mamontov, Burnham, Chen, Moravsky, Loong,
  De~Souza, and Kolesnikov]{@10.1063/1.2194020}
Mamontov,~E.; Burnham,~C.; Chen,~S.-H.; Moravsky,~A.; Loong,~C.-K.;
  De~Souza,~N.; Kolesnikov,~A. Dynamics of water confined in single-and
  double-wall carbon nanotubes. \emph{The Journal of Chemical Physics}
  \textbf{2006}, \emph{124}, 194703\relax
\mciteBstWouldAddEndPuncttrue
\mciteSetBstMidEndSepPunct{\mcitedefaultmidpunct}
{\mcitedefaultendpunct}{\mcitedefaultseppunct}\relax
\EndOfBibitem
\bibitem[Maniwa \latin{et~al.}(2002)Maniwa, Kataura, Abe, Suzuki, Achiba, Kira,
  and Matsuda]{@10.1143/JPSJ.71.2863}
Maniwa,~Y.; Kataura,~H.; Abe,~M.; Suzuki,~S.; Achiba,~Y.; Kira,~H.; Matsuda,~K.
  Phase transition in confined water inside carbon nanotubes. \emph{Journal of
  the Physical Society of Japan} \textbf{2002}, \emph{71}, 2863--2866\relax
\mciteBstWouldAddEndPuncttrue
\mciteSetBstMidEndSepPunct{\mcitedefaultmidpunct}
{\mcitedefaultendpunct}{\mcitedefaultseppunct}\relax
\EndOfBibitem
\bibitem[Maniwa \latin{et~al.}(2005)Maniwa, Kataura, Abe, Udaka, Suzuki,
  Achiba, Kira, Matsuda, Kadowaki, and Okabe]{@10.1016/j.cplett.2004.11.112}
Maniwa,~Y.; Kataura,~H.; Abe,~M.; Udaka,~A.; Suzuki,~S.; Achiba,~Y.; Kira,~H.;
  Matsuda,~K.; Kadowaki,~H.; Okabe,~Y. Ordered water inside carbon nanotubes:
  formation of pentagonal to octagonal ice-nanotubes. \emph{Chemical Physics
  Letters} \textbf{2005}, \emph{401}, 534--538\relax
\mciteBstWouldAddEndPuncttrue
\mciteSetBstMidEndSepPunct{\mcitedefaultmidpunct}
{\mcitedefaultendpunct}{\mcitedefaultseppunct}\relax
\EndOfBibitem
\bibitem[Reiter \latin{et~al.}(2013)Reiter, Deb, Sakurai, Itou, Krishnan, and
  Paddison]{@10.1103/PhysRevLett.111.036803}
Reiter,~G.; Deb,~A.; Sakurai,~Y.; Itou,~M.; Krishnan,~V.; Paddison,~S.
  Anomalous ground state of the electrons in nanoconfined water. \emph{Physical
  Review Letters} \textbf{2013}, \emph{111}, 036803\relax
\mciteBstWouldAddEndPuncttrue
\mciteSetBstMidEndSepPunct{\mcitedefaultmidpunct}
{\mcitedefaultendpunct}{\mcitedefaultseppunct}\relax
\EndOfBibitem
\bibitem[Abou-Hamad \latin{et~al.}(2011)Abou-Hamad, Babaa, Bouhrara, Kim, Saih,
  Dennler, Mauri, Basset, Goze-Bac, and
  W{\aa}gberg]{@10.1103/PhysRevB.84.165417}
Abou-Hamad,~E.; Babaa,~M.-R.; Bouhrara,~M.; Kim,~Y.; Saih,~Y.; Dennler,~S.;
  Mauri,~F.; Basset,~J.-M.; Goze-Bac,~C.; W{\aa}gberg,~T. Structural properties
  of carbon nanotubes derived from 13 C NMR. \emph{Physical Review B}
  \textbf{2011}, \emph{84}, 165417\relax
\mciteBstWouldAddEndPuncttrue
\mciteSetBstMidEndSepPunct{\mcitedefaultmidpunct}
{\mcitedefaultendpunct}{\mcitedefaultseppunct}\relax
\EndOfBibitem
\bibitem[Umeno \latin{et~al.}(2004)Umeno, Kitamura, and
  Kushima]{@10.1016/j.commatsci.2004.02.018}
Umeno,~Y.; Kitamura,~T.; Kushima,~A. Theoretical analysis on electronic
  properties of zigzag-type single-walled carbon nanotubes under radial
  deformation. \emph{Computational Materials Science} \textbf{2004}, \emph{30},
  283--287\relax
\mciteBstWouldAddEndPuncttrue
\mciteSetBstMidEndSepPunct{\mcitedefaultmidpunct}
{\mcitedefaultendpunct}{\mcitedefaultseppunct}\relax
\EndOfBibitem
\bibitem[de~Oliveira \latin{et~al.}(2016)de~Oliveira, Chacham, Soares,
  Manhabosco, de~Resende, and Batista]{@10.1016/j.carbon.2015.09.099}
de~Oliveira,~A.~B.; Chacham,~H.; Soares,~J.~S.; Manhabosco,~T.~M.;
  de~Resende,~H.~F.; Batista,~R.~J. Vibrational G peak splitting in laterally
  functionalized single wall carbon nanotubes: Theory and molecular dynamics
  simulations. \emph{Carbon} \textbf{2016}, \emph{96}, 616--621\relax
\mciteBstWouldAddEndPuncttrue
\mciteSetBstMidEndSepPunct{\mcitedefaultmidpunct}
{\mcitedefaultendpunct}{\mcitedefaultseppunct}\relax
\EndOfBibitem
\bibitem[Vega \latin{et~al.}(2009)Vega, Abascal, Conde, and
  Aragones]{@10.1039/B805531A}
Vega,~C.; Abascal,~J.~L.; Conde,~M.; Aragones,~J. What ice can teach us about
  water interactions: a critical comparison of the performance of different
  water models. \emph{Faraday Discussions} \textbf{2009}, \emph{141},
  251--276\relax
\mciteBstWouldAddEndPuncttrue
\mciteSetBstMidEndSepPunct{\mcitedefaultmidpunct}
{\mcitedefaultendpunct}{\mcitedefaultseppunct}\relax
\EndOfBibitem
\bibitem[Pi \latin{et~al.}(2009)Pi, Aragones, Vega, Noya, Abascal, Gonzalez,
  and McBride]{@10.1080/00268970902784926}
Pi,~H.~L.; Aragones,~J.~L.; Vega,~C.; Noya,~E.~G.; Abascal,~J.~L.;
  Gonzalez,~M.~A.; McBride,~C. Anomalies in water as obtained from computer
  simulations of the TIP4P/2005 model: density maxima, and density, isothermal
  compressibility and heat capacity minima. \emph{Molecular Physics}
  \textbf{2009}, \emph{107}, 365--374\relax
\mciteBstWouldAddEndPuncttrue
\mciteSetBstMidEndSepPunct{\mcitedefaultmidpunct}
{\mcitedefaultendpunct}{\mcitedefaultseppunct}\relax
\EndOfBibitem
\bibitem[Lennard-Jones(1931)]{@10.1088/0959-5309}
Lennard-Jones,~J.~E. Cohesion. \emph{Proceedings of the Physical Society
  (1926-1948)} \textbf{1931}, \emph{43}, 461\relax
\mciteBstWouldAddEndPuncttrue
\mciteSetBstMidEndSepPunct{\mcitedefaultmidpunct}
{\mcitedefaultendpunct}{\mcitedefaultseppunct}\relax
\EndOfBibitem
\bibitem[{Plimpton, Steve}()]{LAMMPS}
{Plimpton, Steve}, {LAMMPS Molecular Dynamics Simulator}.
  \url{https://docs.lammps.org/Manual.html}\relax
\mciteBstWouldAddEndPuncttrue
\mciteSetBstMidEndSepPunct{\mcitedefaultmidpunct}
{\mcitedefaultendpunct}{\mcitedefaultseppunct}\relax
\EndOfBibitem
\bibitem[Plimpton(1995)]{@10.1006/jcph.1995.1039}
Plimpton,~S. Fast parallel algorithms for short-range molecular dynamics.
  \emph{Journal of Computational Physics} \textbf{1995}, \emph{117},
  1--19\relax
\mciteBstWouldAddEndPuncttrue
\mciteSetBstMidEndSepPunct{\mcitedefaultmidpunct}
{\mcitedefaultendpunct}{\mcitedefaultseppunct}\relax
\EndOfBibitem
\bibitem[Ostler \latin{et~al.}(2017)Ostler, Kannam, Daivis, Frascoli, and
  Todd]{@10.1021/acs.jpcc.7b08326}
Ostler,~D.; Kannam,~S.~K.; Daivis,~P.~J.; Frascoli,~F.; Todd,~B. Electropumping
  of water in functionalized carbon nanotubes using rotating electric fields.
  \emph{The Journal of Physical Chemistry C} \textbf{2017}, \emph{121},
  28158--28165\relax
\mciteBstWouldAddEndPuncttrue
\mciteSetBstMidEndSepPunct{\mcitedefaultmidpunct}
{\mcitedefaultendpunct}{\mcitedefaultseppunct}\relax
\EndOfBibitem
\bibitem[Hanasaki and Nakatani(2006)Hanasaki, and Nakatani]{@10.1063/1.2187971}
Hanasaki,~I.; Nakatani,~A. Flow structure of water in carbon nanotubes:
  Poiseuille type or plug-like? \emph{The Journal of Chemical Physics}
  \textbf{2006}, \emph{124}, 144708\relax
\mciteBstWouldAddEndPuncttrue
\mciteSetBstMidEndSepPunct{\mcitedefaultmidpunct}
{\mcitedefaultendpunct}{\mcitedefaultseppunct}\relax
\EndOfBibitem
\bibitem[Kotsalis \latin{et~al.}(2004)Kotsalis, Walther, and
  Koumoutsakos]{@10.1016/j.ijmultiphaseflow.2004.03.009}
Kotsalis,~E.; Walther,~J.~H.; Koumoutsakos,~P. Multiphase water flow inside
  carbon nanotubes. \emph{International Journal of Multiphase Flow}
  \textbf{2004}, \emph{30}, 995--1010\relax
\mciteBstWouldAddEndPuncttrue
\mciteSetBstMidEndSepPunct{\mcitedefaultmidpunct}
{\mcitedefaultendpunct}{\mcitedefaultseppunct}\relax
\EndOfBibitem
\bibitem[Joseph and Aluru(2008)Joseph, and Aluru]{@10.1021/nl072385q}
Joseph,~S.; Aluru,~N. Why are carbon nanotubes fast transporters of water?
  \emph{Nano Letters} \textbf{2008}, \emph{8}, 452--458\relax
\mciteBstWouldAddEndPuncttrue
\mciteSetBstMidEndSepPunct{\mcitedefaultmidpunct}
{\mcitedefaultendpunct}{\mcitedefaultseppunct}\relax
\EndOfBibitem
\bibitem[Harris and Woolf(1980)Harris, and Woolf]{@10.1039/F19807600377}
Harris,~K.~R.; Woolf,~L.~A. Pressure and temperature dependence of the self
  diffusion coefficient of water and oxygen-18 water. \emph{Journal of the
  Chemical Society, Faraday Transactions 1: Physical Chemistry in Condensed
  Phases} \textbf{1980}, \emph{76}, 377--385\relax
\mciteBstWouldAddEndPuncttrue
\mciteSetBstMidEndSepPunct{\mcitedefaultmidpunct}
{\mcitedefaultendpunct}{\mcitedefaultseppunct}\relax
\EndOfBibitem
\bibitem[Raabe and Sadus(2012)Raabe, and Sadus]{@10.1063/1.4749382}
Raabe,~G.; Sadus,~R.~J. Molecular dynamics simulation of the effect of bond
  flexibility on the transport properties of water. \emph{The Journal of
  Chemical Physics} \textbf{2012}, \emph{137}, 104512\relax
\mciteBstWouldAddEndPuncttrue
\mciteSetBstMidEndSepPunct{\mcitedefaultmidpunct}
{\mcitedefaultendpunct}{\mcitedefaultseppunct}\relax
\EndOfBibitem
\bibitem[Mendon{\c{c}}a \latin{et~al.}(2020)Mendon{\c{c}}a, Ternes, Salcedo,
  de~Oliveira, and Barbosa]{@10.1063/5.0031084}
Mendon{\c{c}}a,~B. H.~S.; Ternes,~P.; Salcedo,~E.; de~Oliveira,~A.~B.;
  Barbosa,~M.~C. Water diffusion in carbon nanotubes: Interplay between
  confinement, surface deformation, and temperature. \emph{The Journal of
  Chemical Physics} \textbf{2020}, \emph{153}, 244504\relax
\mciteBstWouldAddEndPuncttrue
\mciteSetBstMidEndSepPunct{\mcitedefaultmidpunct}
{\mcitedefaultendpunct}{\mcitedefaultseppunct}\relax
\EndOfBibitem
\end{mcitethebibliography}

\end{document}